\documentclass[reprint,showpacs,showkeys,amsmath,amssymb,aps,nofootinbib,prd]{revtex4-1}
\usepackage{graphicx}
\usepackage{epstopdf}
\usepackage{amssymb}
\begin{document}
\title{
Randall-Sundrum versus holographic cosmology
}
  \author{Neven Bili\'c}
\email{bilic@irb.hr}
\affiliation{
Division of Theoretical Physics, \\ Rudjer Bo\v{s}kovi\'{c} Institute,
P.O.\ Box 180, 10001 Zagreb, Croatia
}
\date{\today}

\begin{abstract}
We consider a model of a holographic  braneworld universe
in which a cosmological fluid occupies a 3+1-dimensional brane located at 
the boundary of the asymptotic  AdS$_5$ bulk. We  combine the AdS/CFT correspondence 
and the second Randall-Sundrum (RSII) model to establish
a relationship between the RSII braneworld cosmology  and the boundary metric
induced by the time dependent bulk geometry.
In the framework of the Friedmann Robertson Walker cosmology we discuss some physically interesting
scenarios involving the RSII and holographic braneworlds.
\end{abstract}
%
\maketitle

\section{Introduction}
\label{introduction}

The AdS/CFT correspondence  establishes 
 an equivalence of a four-dimensional ${\mathcal{N}}=4$ supersymmetric Yang-Mills theory
and string theory in a ten-dimensional ${\rm AdS}_5\times {\rm S}_5$
bulk \cite{maldacena,gubser,witten1}.
In a wider context of
 gage-gravity duality 
the AdS/CFT correspondence goes beyond pure string theory and  links 
many other important theoretical and phenomenological issues. 
 In particular, a simple physically relevant model related to  AdS/CFT  
is the Randall-Sundrum model \cite{randall1,randall2} and its cosmological applications.
The model was originally proposed as a solution to the hierarchy problem in particle physics
and as a possible mechanism for localizing gravity on the 3+1-dimensional universe embedded in
a 4+1 spacetime without compactification of the extra dimension.
Soon after the papers  \cite{randall1,randall2} appeared, it was realized that the Randall-Sundrum model
is deeply rooted in a wider framework of AdS/CFT correspondence \cite{gubser2,nojiri1, giddings,hawking,duff3,nojiri2,haro2}.
In the braneworld scheme the RS brane provides a cutoff regularization
for the infrared divergences of the on shell bulk action.
 
Our purpose is to study in terms of the  AdS/CFT correspondence a class  of
3+1 time dependent metrics induced on slices of the 4+1-dimensional asymptotic AdS$_5$
bulk.
We consider two types of  braneworld universes:
the holographic braneworld in which  a 3+1-dimensional brane is located at 
the boundary of the
  AdS$_5$ bulk and the Randall-Sundrum (RSII) braneworld in which  a single brane is located at
  a nonzero distance from the boundary.
  We  combine the holographic map of 
Apostolopoulos, Siopsis, and Tetradis \cite{apostolopoulos,tetradis}
and the homogeneous cosmology of the RSII model \cite{randall2}
to establish a mapping between the RSII braneworld cosmology 
and the FRW type cosmology on the holographic braneworld.
We explicitly determine the functional relations between  the two cosmologies
in terms of cosmological scales, Hubble rates and effective densities on the branes.

Our approach is in spirit similar to Brax and Peschanski \cite{brax} but we have included some salient features
which were not sufficiently emphasized in the literature. 
In particular, in connection with the holographic map, we carefully analyze two versions of the RSII models:
the so called ``one-sided'' and ``two-sided'' version.
A general asymptotically AdS metric in Fefferman-Graham coordinates \cite{fefferman}
is of the form
\begin{equation}
ds^2=G_{ab}dx^adx^b =\frac{\ell^2}{z^2}\left( g_{\mu\nu} dx^\mu dx^\nu -dz^2\right),
 \label{eq3001}
\end{equation}
where we use the Latin alphabet for bulk indices
and the Greek alphabet for 3+1 spacetime indices.
In the original  RSII model one assumes
the $Z_2$ symmetry  $z\leftrightarrow z_{\rm br}^2/z$,  so
the region $0<z\leq z_{\rm br}$ is identified with $z_{\rm br} \leq z <\infty$, 
with the observer brane at the fixed point $z=z_{\rm br}$.
Hence, the braneworld is sitting between two patches of AdS$_5$, one on either side, and is
therefore dubbed ``two sided'' \cite{duff3,haro2}.
In contrast, in the ``one-sided'' 
 RSII model the region $0\leq z\leq z_{\rm br}$ is simply cut off 
so the bulk is the section of spacetime $z_{\rm br} \leq z <\infty$.
These two versions are equivalent from the point of view of an observer at the braneworld.
However, in the one-sided RSII model, as pointed out by Duff and Liu \cite{duff3},  
by shifting the boundary in AdS$_5$
from $z=0$ to $z=z_{\rm br}$ 
the model is conjectured to be dual to a cutoff conformal field theory (CFT)
coupled to gravity, with $z=z_{\rm br}$ providing
the cutoff. This conjecture  then reduces to
the standard AdS/CFT duality as the boundary is pushed off to $z=0$. This
connection involves a single CFT at the boundary  of a single patch of AdS$_5$. 
In the  two-sided RSII model  
one would instead require two copies of the CFT, one for each of the AdS$_5$ patches.
We shall demonstrate this explicitly in Sec.\ \ref{map}.
The holographic  mapping turns out to be unique for the two-sided RSII model
whereas in the one-sided model the mapping from the holographic to the RSII cosmology
is a two-valued function. 

The remainder of the paper is organized as follows. 
In Sec.\ \ref{randall} we present a brief derivation of the 
cosmology on the RSII brane. 
In  Sec.\ \ref{holographic} we discuss the cosmology on the holographic brane.
The map from RSII to holographic cosmology is constructed in Sec.\ \ref{map} where we confront   
two cosmological scenarios. We compare the corresponding effective energy densities
and equations of state of the cosmological fluid and discuss a few physically interesting regimes.  
In the concluding section, Sec.\ \ref{conclusion}, we summarize our results and give conclusions.
A brief review of the RSII model is presented in Appendix \ref{rsII} and 
a connection between RSII and AdS/CFT correspondence is demonstrated in Appendix \ref{connection}
where we derive the field equations on the boundary brane  with matter and discuss the conformal anomaly.

\section{Randall-Sundrum cosmology}
\label{randall}

Branewarld cosmology is based on the  scenario in which matter is confined 
on a brane moving in the higher-dimensional bulk
with only gravity allowed to propagate in the bulk  \cite{randall1,randall2,arkani,antoniadis}.
The RS model was originally proposed as a possible mechanism for localizing gravity
on the 3+1 universe embedded in a 4+1-dimensional spacetime without compactification of the extra dimension.
The RSII model is a 4+1-dimensional Anti de Sitter universe containing two 3-branes with opposite tensions
separated in the fifth dimension: observers reside on the positive tension brane 
and the negative tension brane is pushed off to infinity.
The Planck mass scale is determined by the curvature of the AdS spacetime rather than 
by the size of the fifth dimension.
Hence, the model  provides an alternative to compactification \cite{randall2}.

As demonstrated in Appendix \ref{rsII}, in this model 
the fifth dimension can be integrated out to obtain a purely four-dimensional action
with a well-defined value for  Newton's constant in terms of the AdS curvature radius $\ell$ and the 
five-dimensional gravitational constant $G_5$
\begin{equation}
 G_{\rm N} = \frac{2 G_5}{\gamma\ell} ,
 \label{eq0001}
\end{equation}
where we have introduced the {\em sidedness} constant $\gamma$ to facilitate a joint description
of the two versions of the RSII model: 
the one-sided ($\gamma=1$) and two-sided ($\gamma=2$).
In the following  analysis we shall consider $G_{\rm N}$  and $\ell$ as fixed basic 
physical parameters and $G_5$ as a derived quantity.

The classical 3+1-dimensional  gravity on the RSII brane is altered due to the extra dimension.
It has been shown
\cite{garriga} that for $r\gg \ell$ 
the weak gravitational potential  created by an isolated matter source on the brane 
is given by 
\begin{equation} 
\Phi(r)=\frac{G_{\rm N} M}{r}\left( 1+ \frac{2\ell^2}{3r^2} \right) .
\label{eq0018} 
\end{equation}
Hence the extra-dimension  effects 
strengthen Newton's gravitational field.
Table-top tests of Newton's laws \cite{long} currently find no deviations of Newton's potential
at distances greater than 0.1 mm yielding the limit on the AdS$_5$
curvature
\begin{equation} 
\ell < 0.1 {\rm mm}, \quad {\rm or} \quad\ell^{-1} >10^{-12} {\rm GeV}.
\label{eq0021} 
\end{equation}
Assuming (\ref{eq0001}), this yields a lower bound  on the bulk scale parameter \cite{maartens}  
\begin{equation} 
M_5=G_5^{-1/3} >10^8 {\rm GeV} .
\label{eq0019} 
\end{equation}

Soon after Randall and Sundrum introduced their model \cite{randall1,randall2}  it was realized that the model 
as well as any similar braneworld model, 
may have interesting cosmological
implications \cite{binetruy,kraus,flanagan,mukohyama}. In particular,  the usual 
Friedmann equations are modified
so the model has predictions different from the standard cosmology.

To study the braneworld cosmology 
it is convenient to represent the bulk metric 
in  Schwarzschild coordinates \cite{birmingham1}
\begin{equation}
ds_{\rm ASch}^2= 
f(r) dt^2- \frac{dr^2}{f(r)} -r^2 d\Omega_\kappa^2, 
\label{eq3202}
\end{equation}
where 
\begin{equation}
f(r)=\frac{r^2}{\ell^2}+\kappa -\mu \frac{\ell^2}{r^2},
\label{eq3225}
\end{equation}
and
\begin{equation}
d\Omega^2_\kappa=d\chi^2+\frac{\sin^2(\sqrt{\kappa}\chi)}{\kappa}(d\vartheta^2+\sin^2 \vartheta d\varphi^2)
\label{eq1004}
\end{equation}
is the spatial line element for a 
closed ($\kappa=1$), open hyperbolic ($\kappa=-1$), or open flat ($\kappa=0$) space.
The dimensionless parameter $\mu$ is  related to the black-hole mass via \cite{myers,witten2}
\begin{equation}
\mu=\frac{8G_5 M_{\rm bh}}{3\pi \ell^2}.
 \label{eq3105}
\end{equation}
As shown in Appendix \ref{rsII}, for a time dependent brane hypersurface defined by
\begin{equation}
r-a(t)=0,
\label{eq012}
\end{equation}
where $a=a(t)$ is an arbitrary function,
the induced line element on the brane is given by
\begin{equation}
ds_{\rm ind}^2=n^2(t)dt^2 -a(t)^2 d\Omega_\kappa^2 ,
\label{eq015}
\end{equation}
with  the lapse function
\begin{equation}
n^2 =f(a)-\frac{(\partial_t a)^2}{f(a)} .
\label{eq026}
\end{equation}
The effective Friedmann equation on the RSII brane  derived in Appendix \ref{rsII}  
reads
\begin{equation}
\mathcal{H}_{\rm RSII}^2 =\frac{(\sigma+\rho)^2}{\ell^2\sigma_0^2}-\frac{1}{\ell^2} 
+\frac{\mu\ell^2}{a^4} ,
\label{eq032}
\end{equation}
where 
\begin{equation}
 \mathcal{H}_{\rm RSII}^2=H_{\rm RSII}^2+\frac{\kappa}{a^2}=
 \frac{(\partial_t a)^2}{n^2 a^2}+\frac{\kappa}{a^2} .
\label{eq023}
\end{equation}
From now on, a calligraphic $\mathcal{H}$ will always denote 
the Hubble rate $H$ plus the corresponding curvature term $\kappa/a^2$.
The quantity $\sigma$ is the brane tension and we have introduced a constant
\begin{equation} 
\sigma_0= \frac{3\gamma}{8\pi G_5\ell}=\frac{3}{4\pi G_{\rm N}\ell^2} 
\label{eq2004} 
\end{equation}
the value of which is restricted by
\begin{equation} 
\sigma_0 > (10^3 {\rm GeV})^4
\label{eq2006} 
\end{equation}
on account of the experimental constraint (\ref{eq0021}).
Employing the RSII fine-tuning condition $\sigma=\sigma_0$ and (\ref{eq0001}),  Eq.\ (\ref{eq032}) may  be  expressed in the form
\begin{equation}
\mathcal{H}_{\rm RSII}^2 =\frac{8\pi G_{\rm N}}{3}\rho\left(1+ \frac{\rho}{2\sigma_0}\right) 
+\frac{\mu\ell^2}{a^4} ,
\label{eq029}
\end{equation}
which differs from the standard  Friedmann equation 
and is therefore subject to cosmological tests (see, e.g., Refs. \cite{maartens,godlowski}).
The deviation proportional to $\rho^2$  poses no  problem
as it decays as $a^{-8}$ in
the radiation epoch and  will rapidly become negligible after the end of the
high-energy regime $\rho\simeq \sigma_0$ \cite{maartens}.
The last term on the right-hand side of (\ref{eq029}), the so called ``dark radiation'', for positive $\mu$
 should not
exceed 10\% of the total radiation content in
the epoch of big bang (BB) nucleosynthesis  whereas for negative $\mu$ could be as large as
the rest of the radiation content \cite{ichiki,bratt}.
As expected, both the one-sided and two-sided versions of the RSII model yield identical braneworld cosmologies.

Combining the  time derivative of (\ref{eq029})
with the energy conservation  
one finds  the second Friedmann equation (\ref{eq3226}) 
which may be expressed as 
\begin{equation}
 \frac{1}{an}\frac{d}{dt}\left(\frac{1}{n}\frac{da}{dt}\right)+ \mathcal{H}_{\rm RSII}^2=
 \frac{4\pi G_{\rm N}}{3}(\rho-3p)
 -\frac{\rho}{\ell^2\sigma_0^2}(\rho+3p) .
  \label{eq3223}
\end{equation}
Note that the quadratic terms, i.e., the terms proportional 
to  $\rho^2$ and $\rho p$ in  (\ref{eq029}) and (\ref{eq3223})
may be neglected in the low energy limit $\ell \mathcal{H}_{\rm RSII}\ll 1$.
In that limit Eqs.\ (\ref{eq029}) and (\ref{eq3223}) reduce to the standard
Friedmann equations for a two-component fluid consisting of dark radiation and the fluid
obeying the equation of state $p=p(\rho)$.

For the purpose of comparison of the RSII and  holographic cosmologies to be discussed in Sec.\ \ref{map}
it will be convenient to express the Friedman equation in terms of the metric in Fefferman-Graham coordinates 
 for a brane placed at an arbitrary fixed 
$z=z_{\rm br}$.
To this end
we transform the static bulk metric in Schwarzschild 
coordinates $(r,t)$ to the time dependent metric in Fefferman-Graham coordinates $(z,\tau)$ 
in such a way that  the time dependent brane position given by (\ref{eq012})  is fixed at $z=z_{\rm br}$.
Starting from (\ref{eq3202}) we make the coordinate transformation 
\begin{equation}
 t=t(\tau,z), \quad  r=r(\tau,z).
 \label{eq204}
\end{equation}
Then, the line element in new coordinates  will have a general form
\begin{equation}
ds_{(5)}^2=\frac{\ell^2}{z^2}\left( 
{\mathcal{N}}^2(\tau,z)d\tau^2- {\mathcal{A}}^2(\tau,z) d\Omega_\kappa^2-dz^2 
\right) ,
 \label{eq102}
\end{equation}
where 
\begin{equation}
 {\mathcal{A}}^2(\tau,z)=\frac{z^2}{\ell^2}r^2(\tau,z) .  
 \label{eq103}
\end{equation}
To recover
the induced metric (\ref{eq015}) on the brane at $z=z_{\rm br}$
the functions ${\mathcal{A}}$ and  ${\mathcal{N}}$  should satisfy the conditions
\begin{equation}
\frac{\ell^2}{z_{\rm br}^2} \mathcal{A}^2(\tau,z_{\rm br})= a^2(t(\tau,z_{\rm br})), 
\label{eq104}
\end{equation}
\begin{equation}
 \frac{\ell^2}{z_{\rm br}^2} \mathcal{N}^2(\tau,z_{\rm br})=  \dot{t}(\tau,z_{\rm br})^2 n^2(t(\tau,z_{\rm br})) ,
 \label{eq1041}
\end{equation}
where the overdot denotes a derivative with respect to $\tau$.
Besides, from (\ref{eq012}), it follows that
\begin{equation}
 r(\tau,z_{\rm br})=a(t(\tau,z_{\rm br})).
 \label{eq2001}
\end{equation}
Using (\ref{eq104}), the quantity $\mathcal{H}_{\rm RSII}$  may be expressed 
in terms of $\mathcal{A}_{\rm br}(\tau)=\mathcal{A}(\tau,z_{\rm br})$ and $\mathcal{N}_{\rm br}(\tau)=\mathcal{N}(\tau,z_{\rm br})$:
\begin{equation}
 \frac{\ell^2}{z_{\rm br}^2}\mathcal{H}_{\rm RSII}^2=\mathcal{H}_{\rm br}^2
 =\frac{\dot{\mathcal{A}_{\rm br}}^2}{\mathcal{A}_{\rm br}^2\mathcal{N}_{\rm br}^2}+\frac{\kappa}{\mathcal{A}_{\rm br}^2}.
\label{eq024}
\end{equation}
Then, the effective Friedmann equation (\ref{eq032}) on the $z_{\rm br}$-brane
 takes the form 
\begin{equation}
\mathcal{H}_{\rm br}^2 =\frac{(\sigma+\rho)^2}{z_{\rm br}^2\sigma_0^2}-\frac{1}{z_{\rm br}^2} 
+\frac{\mu z_{\rm br}^2}{\mathcal{A}_{\rm br}^4}.
\label{eq033}
\end{equation}
This expression will  be exploited in Sec.\ \ref{map} in the 
mapping between the RSII and holographic cosmologies.

\section{Holographic cosmology}
\label{holographic}

Here we outline a derivation of the Friedmann equations on the holographic brane following
Apostolopoulos et al.\ \cite{apostolopoulos}.
Consider the line element (\ref{eq3001}) for a general asymptotically AdS$_5$ spacetime 
in Fefferman-Graham coordinates. 
The four-dimensional metric  $g_{\mu\nu}$  near the boundary at $z=0$ can be expanded  
as \cite{haro}
\begin{equation}
 g_{\mu\nu}=g^{(0)}_{\mu\nu}+z^2 g^{(2)}_{\mu\nu}+z^4 g^{(4)}_{\mu\nu}+z^6 
g^{(6)}_{\mu\nu}+\ldots \, .
\label{eq3002}
\end{equation}
By plugging this expansion into bulk Einstein's equations (\ref{eq0003}) and
 solving thus obtained equations order by  order in $z$,
the tensors $g^{(n)}_{\mu\nu}$, $n>0$  may be found in terms of the metric $g^{(0)}_{\mu\nu}$
and its curvature tensor $R_{\mu\nu}$.
 The explicit expressions for $g^{(2)}_{\mu\nu}$ and $g^{(4)}_{\mu\nu}$ are found in
 the Appendix A of Ref.\ \cite{haro}. In particular,  we will need 
 \begin{equation}
g^{(2)}_{\mu\nu}=\frac12 \left(R_{\mu\nu}-\frac16 R g^{(0)}_{\mu\nu}\right)
\label{eq3121}
\end{equation}
 and the relation
\begin{equation}
{\rm Tr} g^{(4)}=-\frac14  
{\rm Tr} (g^{(2)})^2 ,  
\label{eq3120}
\end{equation}
where the trace of a tensor $A_{\mu\nu}$ is defined as
\begin{equation}
{\rm Tr} A= A_\mu^\mu=
g^{(0)\mu\nu}A_{\mu\nu}  . 
\label{eq3129}
\end{equation}

We assume now that the time dependent bulk metric is of the 
form (\ref{eq102}) such that
\begin{equation}
\mathcal{N}(\tau,0)=1, \quad {\mathcal{A}}(\tau,0) = a_0(t). 
 \label{eq302}
\end{equation}
The boundary geometry is then described by
a general FRW spacetime metric
\begin{equation}
ds_{(0)}^2=g^{(0)}_{\mu\nu}dx^\mu dx^\nu =d\tau^2 -a_0^2(\tau) d\Omega_\kappa^2 .
 \label{eq3201}
\end{equation}
Using effective Einstein's equations (\ref{eq3006}) derived in Appendix \ref{rsII}
we obtain the holographic Friedmann equation
\begin{equation}
 \frac{\dot{a}_0}{a_0}+\frac{\kappa}{a_0^2}=\frac{8\pi G_{\rm N}}{3}
\left(\gamma \langle T^{\rm CFT}_{00}\rangle +T^{\rm matt}_{00}\right),
 \label{eq4110}
\end{equation}
where 
 $T^{\rm matt}_{\mu\nu}$ is the energy-momentum tensor associated with matter on the holographic brane 
and  $T^{\rm CFT}_{\mu\nu}$ the energy-momentum tensor of
the CFT on the boundary.
According to the AdS/CFT prescription, the expectation value $\langle T^{\rm CFT}_{\mu\nu}\rangle$
is obtained 
by functionally differentiating the renormalized on-shell bulk gravitational action with respect to the
boundary metric $g^{(0)}_{\mu\nu}$. 
With this procedure, referred to as {\em holographic renormalization},  
one finds 
 \cite{haro} 
\begin{eqnarray}
  \langle T^{\rm CFT}_{\mu\nu}\rangle &=& -\frac{\ell^3}{4\pi G_5}\left\{
 g^{(4)}_{\mu\nu}-\frac18 \left[({\rm Tr} g^{(2)})^2-{\rm Tr} (g^{(2)})^2\right]g^{(0)}_{\mu\nu}
 \right.
 \nonumber
\\ 
&& \left.
 -\frac12 (g^{(2)})^2_{\mu\nu}+\frac14 {\rm Tr} g^{(2)}g^{(2)}_{\mu\nu}
  \right\}.
 \label{eq3106}
\end{eqnarray}
This expression is an explicit realization of the AdS/CFT correspondence: the vacuum expectation value of a
boundary CFT operator is obtained solely in terms of the geometrical quantities of the  bulk.
The components of the tensors $g^{(2)}_{\mu\nu}$ and $g^{(4)}_{\mu\nu}$ may be calculated 
either by applying  the explicit expressions from Ref.\ \cite{haro} to the metric
(\ref{eq3002}) or by expanding the metric (\ref{eq102})  near $z=0$ and
comparing the  $z^2$ and $z^4$ terms with the corresponding ones in the expansion (\ref{eq3002}).
Then, from (\ref{eq3106}) one obtains 
\begin{equation}
 \langle T^{\rm CFT}_{\mu\nu}\rangle = t_{\mu\nu}+
\frac14 \langle {T^{\rm CFT}}^\alpha_\alpha\rangle g^{(0)}_{\mu\nu} .
 \label{eq3107}
\end{equation}
The first term on the right-hand side is
 a traceless tensor the nonvanishing components of which are
\begin{equation}
 t_{00}=-3 t^i_i =\frac{3\ell^3}{64\pi G_5 }
\left(\mathcal{H}_0^4 +\frac{4\mu}{a_0^4} -\frac{\ddot{a}_0}{\dot{a}_0}\mathcal{H}_0^2\right),
 \label{eq3108}
\end{equation}
where 
\begin{equation}
 \mathcal{H}_0^2=H_0^2+\frac{\kappa}{a_0^2},
 \label{eq3111}
\end{equation}
and $H_0=\dot{a}_0/a_0$ is the Hubble expansion rate  on  the $z=0$ boundary.
The second term on the right-hand side of (\ref{eq3107}) corresponds to the conformal anomaly 
\begin{equation}
 \langle {T^{\rm CFT}}^\alpha_\alpha\rangle = 
\frac{3\ell^3}{16\pi G_5}\frac{\ddot{a}_0}{a_0}\mathcal{H}_0^2 .
 \label{eq3027}
\end{equation}
Hence, the CFT dual to the time dependent asymptotically  AdS$_5$ bulk metric (\ref{eq102})
is a conformal fluid with the equation  of state $p_{\rm CFT}=\rho_{\rm CFT}/3$,
where $\rho_{\rm CFT}=t_{00}$, $p_{\rm CFT}=-t^i_i$.
In a static case, i.e., when $\dot{a_0}=0$, the fluid is 
 dual to the AdS$_5$ black hole 
with the energy density related to the black-hole  mass $M_{\rm bh}$ defined in (\ref{eq3105}) 
as 
\begin{equation}
\rho_{\rm CFT}=\frac{M_{\rm bh}}{V} + \frac{3\kappa^2}{64\pi G_5\ell} , 
\end{equation}
where $V=2\pi^2\ell^3$ is the volume of the three-dimensional space 
for a spherical geometry. 
If the boundary geometry is FRW, the dual conformal fluid 
behaves as radiation, the so called "dark radiation".

So far in our consideration the cosmological scale $a_0(\tau)$ at the boundary is assumed 
to be an arbitrary function of $\tau$. In order to satisfy the appropriate boundary condition
for a given $a_0(\tau)$ we place a brane on the boundary with matter  described by the energy-momentum tensor
\begin{equation}
  T^{\rm matt}_{00}= \rho_0 , \quad T^{\rm matt}_{ij}= p_0 g^{(0)}_{ij} ,
 \label{eq3109}
\end{equation}
where $\rho_0$ and $p_0$ are  the total density and pressure,  respectively, 
including the brane tension $\sigma_{\rm br}$
\begin{equation}
\rho_0=\rho_{\rm matt}+\sigma_{\rm br}, \quad p_0=p_{\rm matt}-\sigma_{\rm br}.
 \label{eq3213}
\end{equation}
The Einstein equations (\ref{eq3006})
together with (\ref{eq3109}), (\ref{eq3107}), and (\ref{eq3108})
 yield the holographic Friedmann equation \cite{apostolopoulos,kiritsis}
\begin{equation}
 \mathcal{H}_0^2=\frac{\ell^2}{4}
 \left(\mathcal{H}_0^4+ \frac{4\mu}{a_0^4} \right)+\frac{8\pi G_{\rm N}}{3}\rho_0.
 \label{eq3110}
\end{equation}
Note that the coefficient of the quartic term does not depend on whether 
one is using a one-sided or a two-sided regularization.
Equation (\ref{eq3110}) was derived by Kiritsis \cite{kiritsis}
and independently by Apostolopoulos et al.\ \cite{apostolopoulos} albeit they disagree
in the coefficient of the quartic term\footnote{The reason for 
the disagreement is twofold: first, there is a difference by a factor of 2
because the regularization used in Ref.\ \cite{apostolopoulos} was one sided
whereas in Ref.\ \cite{kiritsis} was two sided.
Another factor of 2 disagreement is due to an unconventional
definition of the stress tensor in Ref. \cite{kiritsis}.}.
Solving the quadratic equation (\ref{eq3110})
one finds  $\mathcal{H}_0$ expressed as an explicit function of $\rho_0$:
\begin{equation}
 \mathcal{H}_0^2=\frac{2}{\ell^2}\left(
 1+\epsilon \sqrt{1-\frac{2\rho_0}{\sigma_0}-\frac{\mu\ell^4}{a_0^4}} \right) ,
 \label{eq3115}
\end{equation}
where $\epsilon=+1$ or $-1$ and
$\sigma_0$ is a constant defined in (\ref{eq2004}). 

For $\epsilon=-1$ the physical range of
the expansion parameter $\mathcal{H}_0$
 is given by
 \begin{equation}
 0\leq \mathcal{H}_0^2\ell^2 \leq 2 ,
 \label{eq3402}
\end{equation}
 corresponding to the energy density interval
\begin{equation}
 - \frac{\sigma_0}{2} \frac{\mu\ell^4}{a_0^4} \leq \rho_0 \leq \frac{\sigma_0}{2}\left(1- \frac{\mu\ell^4}{a_0^4}\right) .
 \label{eq3400}
\end{equation}
In this case  Eq.\ (\ref{eq3115}) agrees with the RSII Friedmann equation (\ref{eq029})
at quadratic order in $\rho$ and linear order in $\mu$.

For $\epsilon=+1$ the physical range of $\mathcal{H}_0$ is given by 
\begin{equation}
\infty> \mathcal{H}_0^2\ell^2\geq2  ,
 \label{eq3404}
\end{equation}
 corresponding to
\begin{equation}
 - \infty < \rho_0 \leq \frac{\sigma_0}{2}\left(1- \frac{\mu\ell^4}{a_0^4}\right).
 \label{eq3403}
\end{equation}
Note that the density $\rho_0$ is negative when $\mathcal{H}_0^2$ 
lies outside the interval 
 \begin{equation}
2-2\sqrt{1-\mu\ell^4/a_0^4}\leq \mathcal{H}_0^2\ell^2\leq 2+2\sqrt{1-\mu\ell^4/a_0^4}  .
 \label{eq4404}
\end{equation}

The second Friedmann equation is obtained by combining the time derivative of (\ref{eq3110})
with the energy conservation 
\begin{equation}
\dot{\rho}_0+3H_0(\rho_0+p_0)=0 .
 \label{3200}
\end{equation}
One finds
\begin{equation}
 \dot{H}_0-\frac{\kappa}{a_0^2}=-4\pi G_{\rm N}(\rho_0+p_0)+
 \frac{\ell^2}{2}
 \left(\dot{H}_0-\frac{\kappa}{a_0^2}\right)\mathcal{H}_0^2- \frac{2\ell^2\mu}{a_0^4},
\label{eq3112}
\end{equation}
which may also be written in the form
\begin{equation}
 \frac{\ddot{a}_0}{a_0}
 \left(1-\frac{\ell^2}{2}\mathcal{H}_0^2\right)+\mathcal{H}_0^2=
 \frac{4\pi G_{\rm N}}{3}(\rho_0-3p_0).
 \label{eq3113}
\end{equation}
Given $a_0(\tau)$, the Friedmann equations (\ref{eq3110})
and (\ref{eq3112})
on  the boundary describe the equation of state $p_0=p_0(\rho_0)$ 
in a parametric form.

{\em Nota bene (N.B.)}:
As in the RSII cosmology, 
in the low energy limit $\ell \mathcal{H}_0 \ll 1$
Eqs.\ (\ref{eq3110}) and (\ref{eq3113}) reduce to the standard
Friedmann equations for a two component fluid consisting of dark radiation and the fluid
obeying the equation of state $p_0=p_0(\rho_0)$.

Remarkably, Eq.\ (\ref{eq3110}) has been also derived in  other contexts.
For $\kappa=1$ and constant $\rho_0$ with (\ref{eq3127}) 
Eq.\ (\ref{eq3110}) coincides with the saddle point of the 
spatially closed mini superspace partition
function dominated by matter fields conformally coupled 
to gravity \cite{barvinsky}.
A variant of Eq.\ (\ref{eq3110}) has been  derived by Lidsey \cite{lidsey} 
from the generalized uncertainty principle and the first law of thermodynamics 
applied to the apparent horizon entropy. The quartic term with $\kappa=0$ in (\ref{eq3110}) 
has been derived quite recently as a quantum correction to the Friedmann equation
 using thermodynamic arguments at the apparent horizon \cite{viaggiu}.

It is worth addressing the holographic cosmology of de Sitter type,
 i.e., for a constant $\rho_0=\Lambda/(8\pi G_{\rm N})$, with $\mu=0$.
 A static representation of the de Sitter boundary spacetime has been recently
 discussed \cite{fischler} in the context of AdS/CFT.
Using the standard $\kappa=1,0$,  and $-1$ representations
of the de Sitter geometry
\begin{equation}
ds^2\!\!=\!\!\left\{ \begin{array}{ll}
d\tau^2-h^{-2}\cosh^2 h\tau \left(d\chi^2+ \sin^2 \chi d\Omega^2\right), & \!\!\mbox{$\kappa = 1$},\\
d\tau^2-e^{2h\tau} \left(d\chi^2+\chi^2d\Omega^2\right), & \!\!\mbox{$\kappa = 0$},\\
d\tau^2-h^{-2}\sinh^2 h\tau \left(d\chi^2+\sinh^2 \chi d\Omega^2\right), 
& \!\!\mbox{$\kappa = -1$},\end{array} \right.\
\label{eq2002}
\end{equation}
Eq.\ (\ref{eq3115}) yields
\begin{equation}
h^2=\frac{2}{\ell^2}\left(
 1+\epsilon \sqrt{1-\frac{\Lambda}{4\pi G_{\rm N} \sigma_0}} \right).
 \label{eq3117}
\end{equation} 
By making use of (\ref{eq0001}) with $\gamma=1$ and (\ref{eq3127}), Eq.\ (\ref{eq3115}) may be expressed as
\begin{equation}
\mathcal{H}_0^2  =\frac{1}{32\pi b G_{\rm N}}\left(
 1+\epsilon \sqrt{1-\frac{64\pi}{3}b G_{\rm N} \Lambda} \right),
 \label{eq3116}
\end{equation}
which coincides with 
the equation for $\mathcal{H}_0^2$ of Pelinson, Shapiro, and Takakura \cite{pelinson}
for the anomaly induced inflation.  Equation (\ref{eq3116}) with $\epsilon=+1$ and $\Lambda=0$ describes
the Starobinski inflation model \cite{starobinski}. With $\epsilon=-1$ and $\Lambda\ll 1/G_{\rm N}$ one recovers
at linear order the standard de Sitter cosmology with the expansion rate $\mathcal{H}_0^2=\Lambda/3$.

\section{Holographic map}  
\label{map}
The bulk metric  that approaches the  metric (\ref{eq3201}) as we approach the boundary $z=0$
is expressed in the form 
(\ref{eq102}) where the functions $\mathcal{A}$ and $\mathcal{N}$ 
are derived in Ref.\ \cite{apostolopoulos} and are expressed in terms of $a_0$ as 
\begin{equation}
\mathcal{A}^2=a_0^2\left[
\left(1-\frac{\mathcal{H}_0^2 z^2}{4}\right)^2
+ \frac14 \frac{\mu z^4}{a_0^4}
\right],
 \label{eq3103}
\end{equation}
\begin{equation}
{\mathcal{N}}=\frac{\dot{\mathcal{A}}}{\dot{a}_0}.
 \label{eq3104}
\end{equation}
The spacetime (\ref{eq102}) may be regarded as
a $z$ foliation of the bulk with an FRW cosmology on each $z$ slice.
For a constant $a_0$, e.g., $a_0=\ell$, one recovers the static AdS-Schwarzschild solution 
(\ref{eq3202})
in which case the metric at the boundary 
$z=0$ ($r\rightarrow \infty$) represents the static Einstein universe.

The Hubble expansion rate corresponding to a $z$-cosmology is defined as
\begin{equation}
H\equiv\frac{\dot{\mathcal{A}}}{{\mathcal{N}}{\mathcal{A}}}= H_0\frac{a_0}{\mathcal{A}}
 \label{eq3204}
\end{equation}
and similarly
\begin{equation}
\mathcal{H}\equiv H^2+\frac{\kappa}{\mathcal{A}^2}  \mathcal{H}_0\frac{a_0}{\mathcal{A}} ,
 \label{eq201}
\end{equation}
where $\mathcal{H}_0$ is defined by (\ref{eq3111}).

It is of interest to express the cosmological scale ${\mathcal{A}}={\mathcal{A}}(\tau,z)$, the lapse function
${\mathcal{N}}={\mathcal{N}}(\tau,z)$ and the Hubble rate $H=H(\tau,z)$
at an arbitrary $z$ slice in terms of $\mathcal{A}_{\rm br}=\mathcal{A}(\tau,z_{\rm br})$,
$\mathcal{N}_{\rm br}=\mathcal{N}(\tau,z_{\rm br})$,
 and $H_{\rm br}=H(\tau,z_{\rm br})$ on another slice $z_{\rm br}$.
 To make a connection with  the RSII cosmology
  we can identify $(\ell/z_{\rm br})\mathcal{A}_{\rm br} =a(t(\tau,z_{\rm br}))$ and  
  $(\ell/z_{\rm br})\mathcal{N}_{\rm br}= n(t(\tau,z_{\rm br}))$
  (see Appendix \ref{rsII}),
where $a(t)$ and $n(t)$ are the functions that appear in the line element (\ref{eq015}) induced on the RSII brane.

\begin{figure}[t]
\begin{center}
\includegraphics[width=0.6\textwidth,trim= 4cm 24cm 0 0cm]{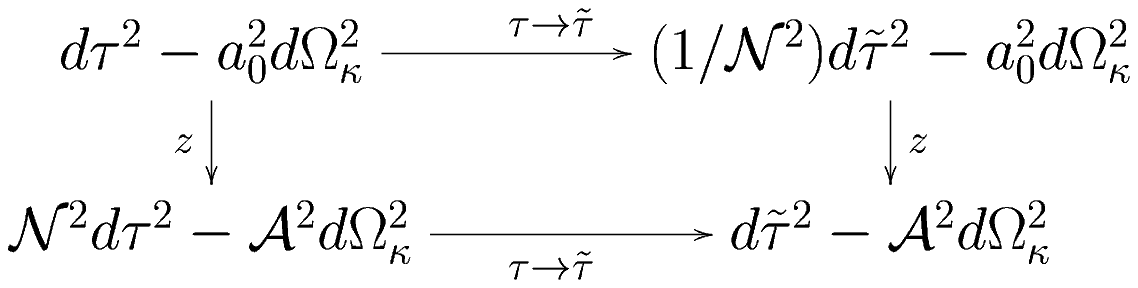}
\caption{Mapping of the  holographic cosmology on the $z=0$ boundary  into the braneworld cosmology on  an arbitrary $z$ slice. 
The times $\tau$ and $\tilde{\tau}$ are the holographic and RSII synchronous times, respectively.
}
\label{fig1}
\end{center}
\end{figure}

First, using (\ref{eq3204}) we can express (\ref{eq3103}) as an equation for $a_0^2$, ${\mathcal{A}}^2$, and ${H^2}$, 
and similarly as another equation for $a_0^2$, $\mathcal{A}_{\rm br}^2$, and $H_{\rm br}^2$. 
Eliminating $a_0^2$ from these
two equations we find
\begin{eqnarray}
\mathcal{A}&=&\frac{\mathcal{A}_{\rm br}}{\sqrt2}
\left[\left(1 +\frac12 \mathcal{H}_{\rm br}^2 z_{\rm br}^2\right)\left(1+\frac{z^4}{z_{\rm br}^4}  \right)
- \mathcal{H}_{\rm br}^2 z^2 
\right. 
\nonumber \\
&&
\left. 
+\epsilon  
\sqrt{1+\mathcal{H}_{\rm br}^2 z_{\rm br}^2
-\frac{\mu z_{\rm br}^4}{\mathcal{A}_{\rm br}^4}}\left(1-\frac{z^4}{z_{\rm br}^4}  \right)
\right]^{1/2} ,
 \label{eq1101}
\end{eqnarray}
where
\begin{equation}
\mathcal{H}_{\rm br}^2=H_{\rm br}^2 +\frac{\kappa}{\mathcal{A}_{\rm br}^2} 
 \label{eq3208}
\end{equation}
and $\epsilon=$ $+1$ or $-1$. 
Thus, the map is not unique due to the sign
ambiguity in front of the square root.
However, consistency with (\ref{eq3103}) in the limit $z_{\rm br}\rightarrow 0$
 and continuity of the metric requires $\epsilon =+1$ in the
region $z\geq z_{\rm br}$, whereas for $z < z_{\rm br}$ both signs are allowed.   
This remaining nonuniqueness  is removed for the two-sided braneworld by the 
$Z_2$ symmetry $z\leftrightarrow z_{\rm br}^2/z$
in which case $\epsilon=-1$  is fixed for the branch 
$z < z_{\rm br}$. Then, the metric (\ref{eq102}) becomes invariant under the transformation
$z\rightarrow \bar{z}=z_{\rm br}^2/z$.
For the benefit of a joint description of one-sided and two-sided versions, the expression (\ref{eq1101}) may be written as 
\begin{eqnarray}
\mathcal{A}&=&\frac{\mathcal{A}_{\rm br}}{\sqrt2}
\left[\left(1 +\frac12 \mathcal{H}_{\rm br}^2 z_{\rm br}^2\right)\left(1+\frac{z^4}{z_{\rm br}^4}  \right)
- \mathcal{H}_{\rm br}^2 z^2 
\right. 
\nonumber \\
&&
\left. 
+\mathcal{E} (z) 
\sqrt{1+\mathcal{H}_{\rm br}^2 z_{\rm br}^2
-\frac{\mu z_{\rm br}^4}{\mathcal{A}_{\rm br}^4}}\left(1-\frac{z^4}{z_{\rm br}^4}  \right)
\right]^{1/2} ,
 \label{eq3205}
\end{eqnarray}
where we have introduced a two-valued step function
\begin{equation}
\mathcal{E}(z)=\left\{ \begin{array}{ll}
+1, & \mbox{ for $z\geq z_{\rm br}$},\\
-1,& \mbox{ for $z<z_{\rm br}$, two-sided version},\\
+1 \mbox{ or } -1, & \mbox{ for $z<z_{\rm br}$, one-sided version} . \end{array} \right.
\label{eq4105}
\end{equation}
Furthermore, applying the definition (\ref{eq3104}) to $\mathcal{N}$ and $\mathcal{N}_{\rm br}$
combined with (\ref{eq3205}), we  find 
\begin{widetext}
\begin{eqnarray}
\frac{\mathcal{N}}{\mathcal{N}_{\rm br}}=\frac{\mathcal{A}}{\mathcal{A}_{\rm br}}+
\left(\frac{z_{\rm br}^2\mathcal{A}_{\rm br}^2\mathcal{H}_{\rm br}\dot{\mathcal{H}}_{\rm br}}{4\mathcal{A}\dot{\mathcal{A}_{\rm br}}}+
\frac{\mu z_{\rm br}^4}{2\mathcal{A}\mathcal{A}_{\rm br}^3}\right)
\frac{\mathcal{E}(z)\left(1-z^4/z_{\rm br}^4  \right)}{\sqrt{1+\mathcal{H}_{\rm br}^2 z_{\rm br}^2
-\mu z_{\rm br}^4/\mathcal{A}_{\rm br}^4 }}
+\frac{z_{\rm br}^2\mathcal{A}_{\rm br}^2\mathcal{H}_{\rm br}\dot{\mathcal{H}}_{\rm br}}{4\mathcal{A}\dot{\mathcal{A}_{\rm br}}}
\left(1-\frac{z^2}{z_{\rm br}^2}\right)^2 .
 \label{eq3305}
\end{eqnarray}
\end{widetext}
The map between the holographic and 
RSII cosmologies is schematically illustrated in Fig.\ \ref{fig1}. 

Note that the quantity $\mathcal{H}_{\rm br}$ in (\ref{eq3205}) and (\ref{eq3305})  
is identical to that defined  in (\ref{eq024}) for the RSII cosmology.
Besides, it is clear by construction that the functions $\mathcal{A}$ and $\mathcal{N}$ in (\ref{eq3205}) and (\ref{eq3305}) 
are, up to a sign, equal to those in the line element (\ref{eq102}).
The general expression (\ref{eq3205})
 agrees with that of Brax and Peschanski \cite{brax} obtained for the two-sided model with $z_{\rm br}=\ell$ and $\kappa=0$.

 \begin{figure*}[ht]
\begin{center}
\includegraphics[width=0.45\textwidth,trim= 0 0 0 0]{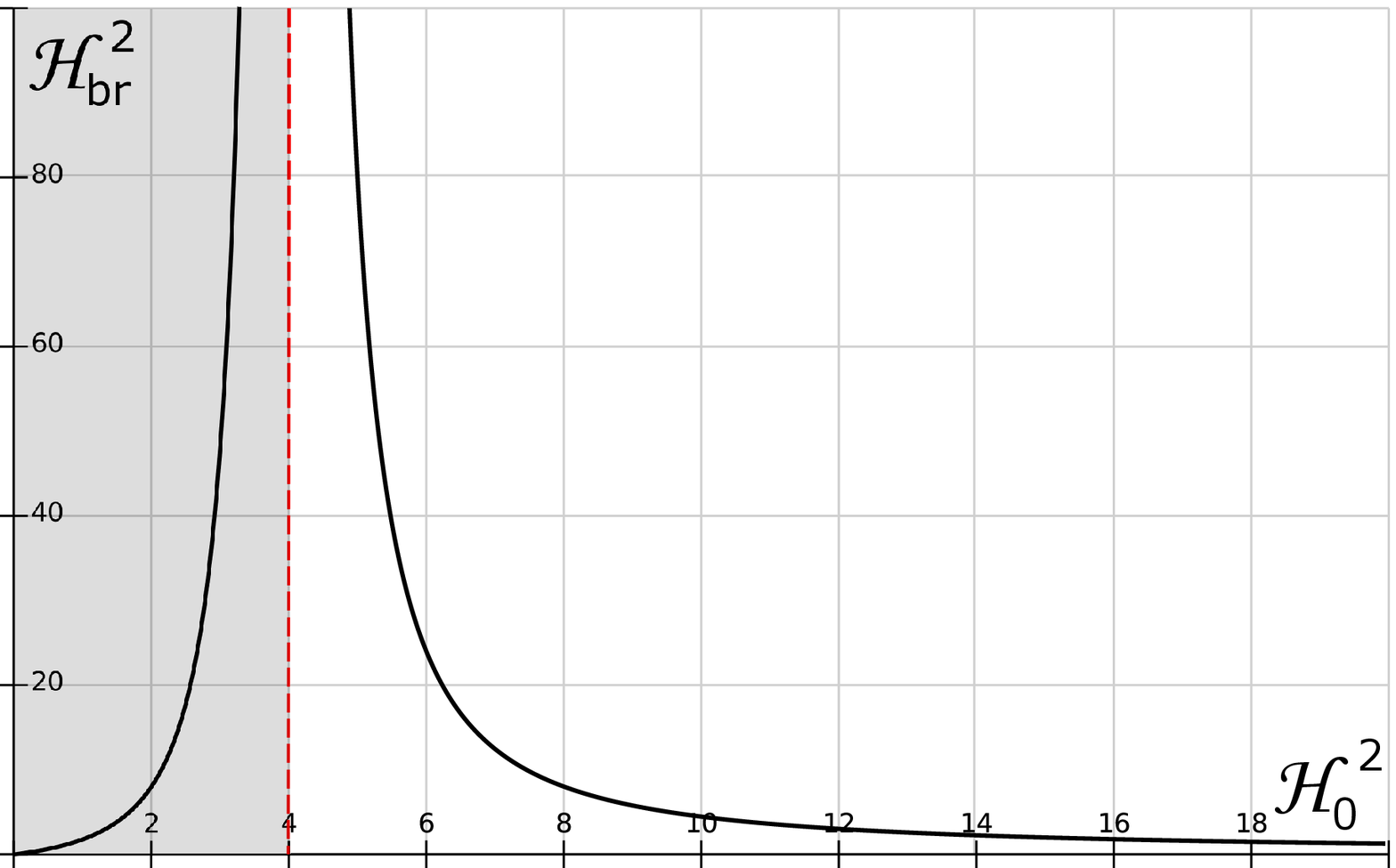}
\hspace{0.02\textwidth}
\includegraphics[width=0.45\textwidth,trim= 0 0 0 0]{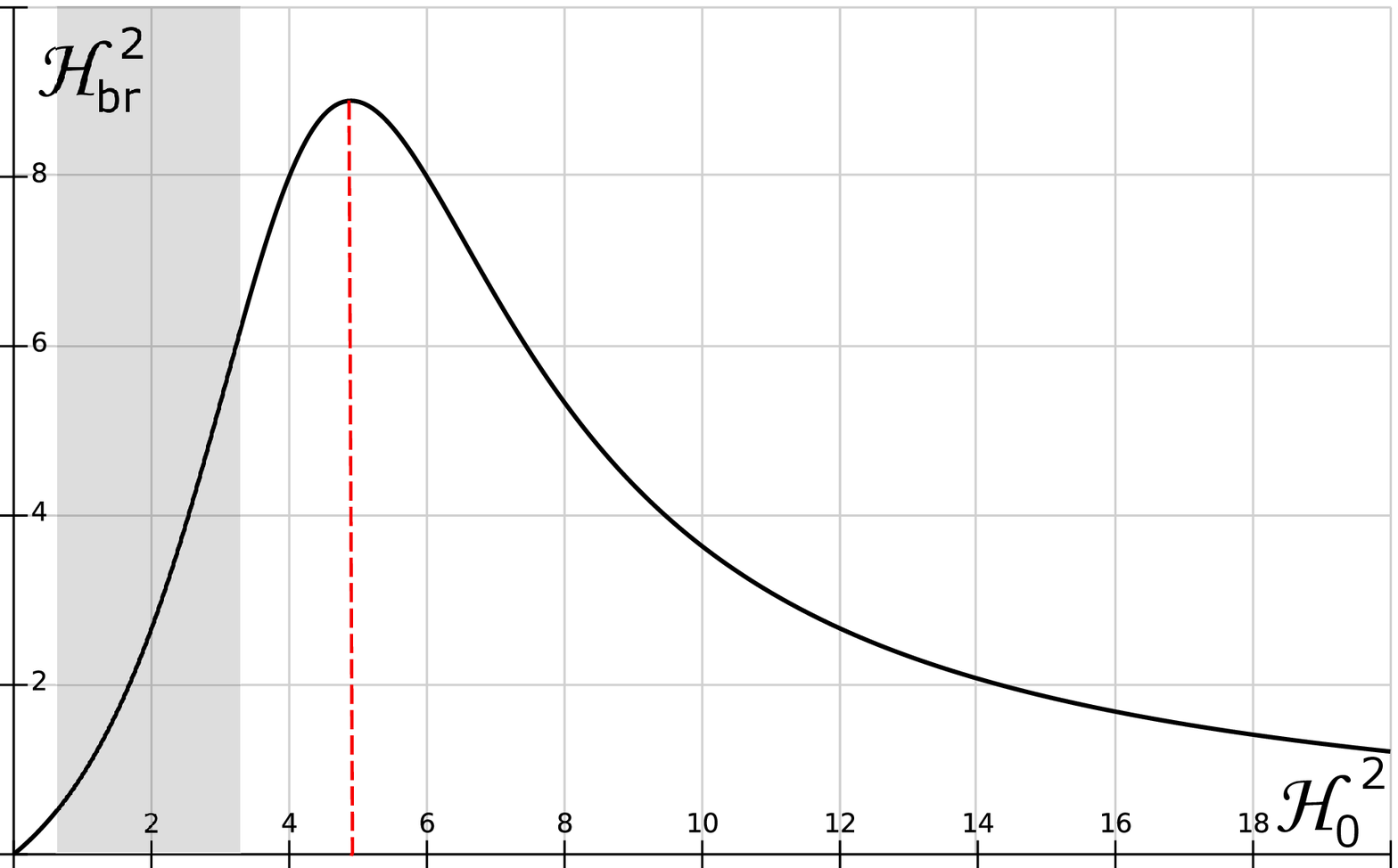}
\caption{$\mathcal{H}_{\rm br}^2$ as a function of $\mathcal{H}_0^2$
(both in units of $z_{\rm br}^{-2}$)  defined by (\ref{eq3211}) for $\mu= 0$ (left panel)
and  $\mu \ell^4/a_0^4=2$ with $z_{\rm br}^2/\ell^2=2$ (right panel).
The region left of the vertical dashed red line is relevant for the one-sided version only.
The shaded area corresponds to the physical region $\rho_0 >0$.
}
\label{fig2}
\end{center}
\end{figure*}

Next we derive a relation  between the Hubble rate $H$  on  an arbitrary $z$ slice and the Hubble rate $H_0$
on the $z=0$ boundary.
Using (\ref{eq3204}) and (\ref{eq201}) we find
\begin{eqnarray}
\mathcal{H}^2&=& 2\mathcal{H}_{\rm br}^2
\left[\left(1 +\frac{\mathcal{H}_{\rm br}^2 z_{\rm br}^2}{2}\right)
\left(1+\frac{z^4}{z_{\rm br}^4}  \right)
- \mathcal{H}_{\rm br}^2 z^2
\right.
\nonumber \\
&&\left.
+\mathcal{E}(z)\sqrt{1+\mathcal{H}_{\rm br}^2 z_{\rm br}^2
-\frac{\mu z_{\rm br}^4}{\mathcal{A}_{\rm br}^4}}\left(1-\frac{z^4}{z_{\rm br}^4}  \right)
\right]^{-1} .
 \label{eq3207}
\end{eqnarray}
Evaluating this  expression  at $z=0$ we find the relationship between
$z=0$ cosmology and the cosmology on  the brane at $z=z_{\rm br}$
\begin{equation}
\mathcal{H}_0^2= 2\mathcal{H}_{\rm br}^2
\left(
1 +\frac{\mathcal{H}_{\rm br}^2 z_{\rm br}^2}{2} +\mathcal{E}_0
\sqrt{1+\mathcal{H}_{\rm br}^2 z_{\rm br}^2
-\frac{\mu z_{\rm br}^4}{\mathcal{A}_{\rm br}^4}}
\right)^{-1} ,
 \label{eq3209}
\end{equation}
where $\mathcal{E}_0\equiv \mathcal{E}(0)=-1$ for the two-sided and $\mathcal{E}_0=+1$ or $-1$ for the one-sided version
of the RSII model.
The inverse relation can be obtained either from (\ref{eq3207}) by taking the limit $z_{\rm br}\rightarrow 0$
and replacing $z\rightarrow z_{\rm br}$
or by making use of (\ref{eq3204}) and (\ref{eq3103}). 
Either way we find
\begin{equation}
\mathcal{H}_{\rm br}^2= \mathcal{H}_0^2\left[1-\frac{\mathcal{H}_0^2 z_{\rm br}^2}{2}
+ \frac{1}{16}\left(\mathcal{H}_0^4+\frac{4\mu}{a_0^4}\right)z_{\rm br}^4 
\right]^{-1}.
 \label{eq3211}
\end{equation}
The functional dependence of $\mathcal{H}_{\rm br}^2$ vs $\mathcal{H}_0^2$
 is depicted in Fig.\ \ref{fig2} for two values of the black-hole mass parameter:
$\mu=0$ (left panel) and $\mu \ell^4/a_0^4=1/2$ with $z_{\rm br}^2/\ell^4=2$ (right panel). 
The shaded areas in both panels denote the region defined by (\ref{eq4404}), i.e., the region
in which $\rho_0 >0$. 
The function assumes a maximal value 
\begin{equation}
\mathcal{H}_{\rm br}^2|_{\rm max}=\frac{8\sqrt{4+\mu z_{\rm br}^4/a_0^4}}{z_{\rm br}^2(\sqrt{4+\mu z_{\rm br}^4/a_0^4}-2)^2} 
 \label{eq4000}
\end{equation}
at
\begin{equation}
\mathcal{H}_0^2|_{\rm max}=\frac{2}{z_{\rm br}^2}\sqrt{4+\frac{\mu z_{\rm br}^4}{a_0^4}}.
 \label{eq4007}
\end{equation}
For $\mu=0$ the maximum becomes a singularity at 
$\mathcal{H}_0^2|_{\rm max}=4/z_{\rm br}^2$.
The part of the domain
where
\begin{equation}
\mathcal{H}_0^2 < \mathcal{H}_0^2|_{\rm max}
 \label{eq4008}
\end{equation}
corresponds to the branch $\mathcal{E}(z)=+1$ for $z<z_{\rm br}$  of Eq.\ (\ref{eq3205}),
and hence the condition (\ref{eq4008}) is met for the one-sided version
only. The remaining part  
\begin{equation}
\mathcal{H}_0^2\geq \mathcal{H}_0^2|_{\rm max} 
 \label{eq4009}
\end{equation}
corresponds to the branch $\mathcal{E}(z)=-1$ for $z<z_{\rm br}$
and is relevant for both the one-sided and two-sided versions.
From (\ref{eq3211}) in the limit $\mathcal{H}_0 \rightarrow \infty$ we find 
\begin{equation}
\mathcal{H}_{\rm br}=\frac{4 }{z_{\rm br}^{2}\mathcal{H}_0}.
 \label{eq3224}
\end{equation}
However, it is important to note that the regime 
in which $\mathcal{H}_0$ does not satisfy Eq. (\ref{eq4404}) violates the weak energy condition
and a large $\mathcal{H}_0$ implies a large negative energy density.
Thus, the large (negative) density limit on the holographic brane
maps into the low -density limit on the RSII brane. 
\begin{figure*}[ht]
\begin{center}
\includegraphics[width=0.45\textwidth ]{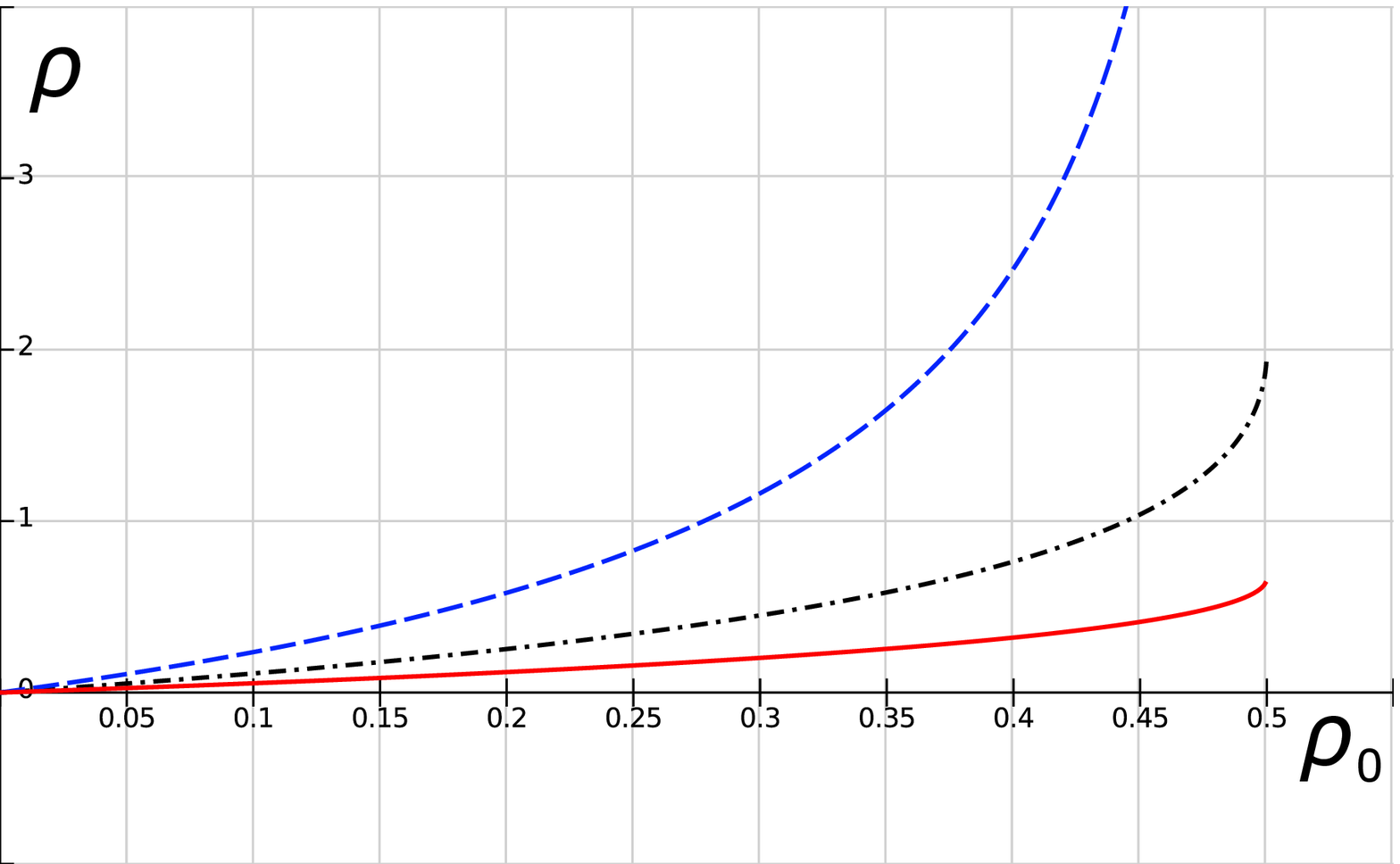}
\hspace{0.02\textwidth}
\includegraphics[width=0.45\textwidth ]{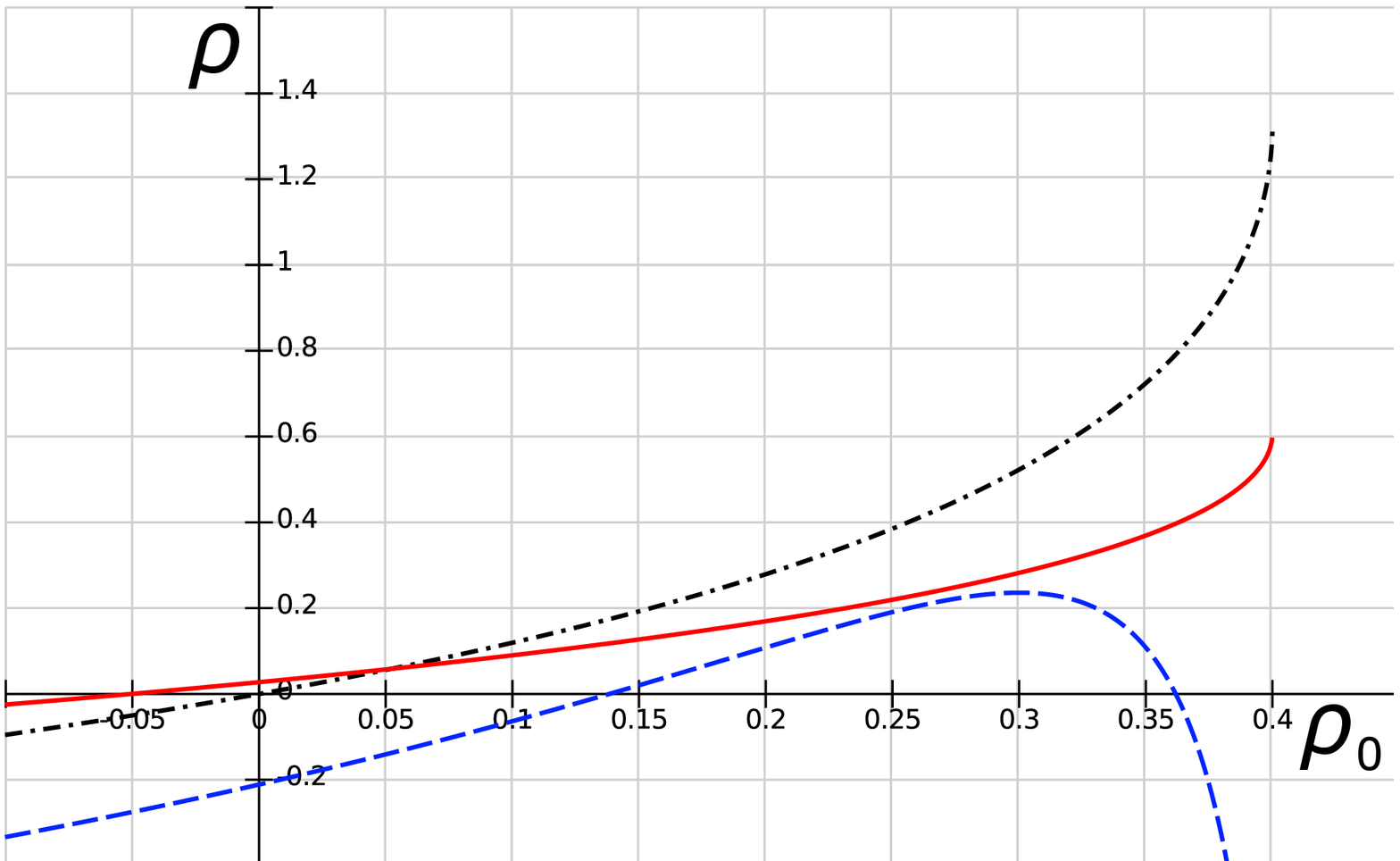}
\\
\vspace{0.02\textwidth}
\includegraphics[width=0.45\textwidth ]{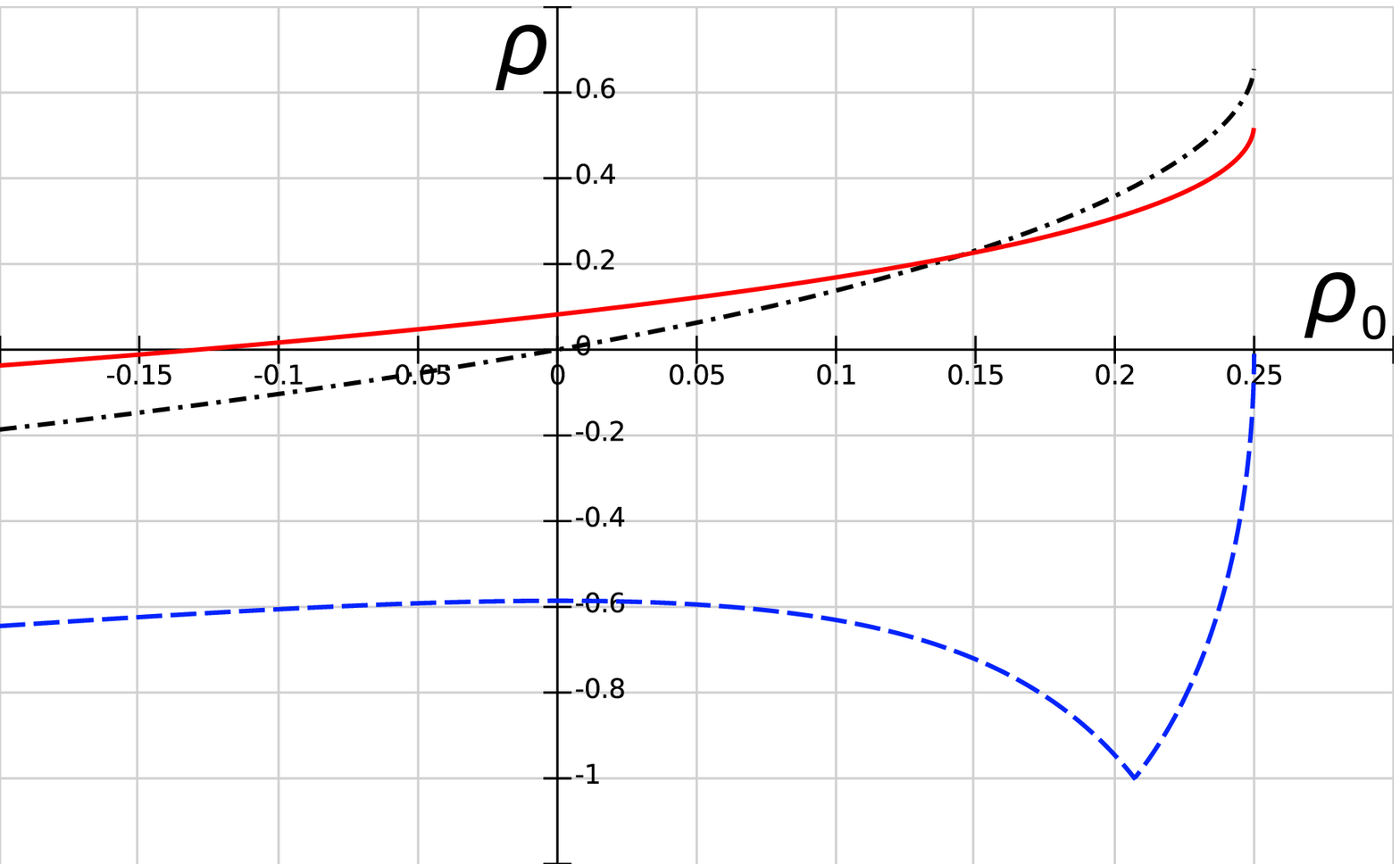}
\hspace{0.02\textwidth}
\includegraphics[width=0.45\textwidth ]{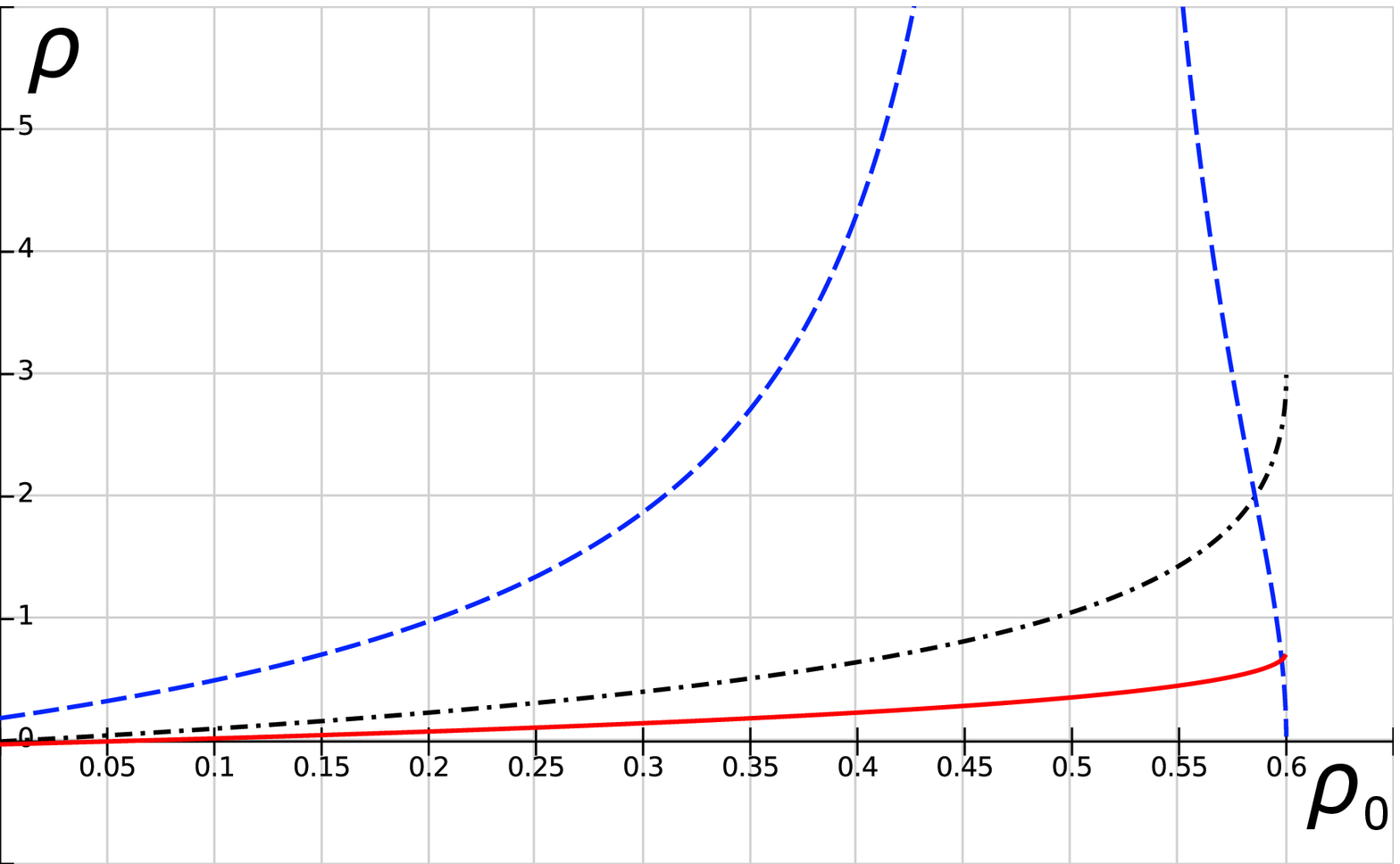}
\caption{The effective density on the RSII brane $\rho$ as a function of the density on the holographic brane $\rho_0$
(both in units of $\sigma_0$) for $\epsilon=-1$, $\sigma=\sigma_0$,
and $\mu \ell^4/a_0^4=$ 0 (top left), 0.2 (top right), 0.5 (bottom left) and $-0.2$ (bottom right).
The full red, dash-dotted black, and dashed blue lines represent $z_{\rm br}^2/\ell^2=$ 0.5, 1,  and 2, respectively.
}
\label{fig3}
\end{center}
\end{figure*}
 
The relationship (\ref{eq3211}) simplifies at a particular point $z_{\rm br}=\sqrt2 \ell$.
Applying (\ref{eq3110}) at $z_{\rm br}=\sqrt2 \ell$
we obtain 
\begin{equation}
\mathcal{H}_{\sqrt2 \ell}^2=\mathcal{H}_0^2
\left(1-\frac{8\pi G_{\rm N} \ell^2}{3}\rho_0\right)^{-1} ,
 \label{eq3212}
\end{equation}
where $\rho_0$ is the effective energy density of matter on the holographic brane,
as defined in (\ref{eq3213}).

{\em N.B.}:
Due to the $Z_2$ symmetry, the  brane at $z=0$ ($y=-\infty$) must be 
identical to the brane at $z=\infty$. 
In the RSII model the second brane is pushed off to $z=\infty$
and hence, the holographic brane at $z=0$ is identical to the second  brane of the RSII model.

Next we analyze a few special cases in two scenarios: the holographic and the RSII cosmological scenario
with the primary braneworld located at $z=0$ and $z=z_{\rm br}$, respectively. 
In each of the two scenarios we assume the presence of matter on the primary brane only and  no matter in the bulk.

\subsection{Holographic scenario}
\label{holoscen}
In the holographic scenario the primary braneworld is at the AdS boundary at $z=0$
evolving according to the Friedmann equations  (\ref{eq3110}) and (\ref{eq3112}).
The cosmology on the RSII brane at an arbitrary $z$ slice  emerges as a  
reflection of the boundary cosmology. 
We would like to express the cosmological parameters on the RSII brane
 at $z=z_{\rm br}$ in terms of the parameters on the holographic brane at $z=0$.
If the density $\rho_0$ and pressure $p_0$ on the holographic brane are known, the
cosmological scale $a_0$ may be derived by integrating (\ref{eq3110}) and (\ref{eq3113}).
On the other hand, given $a_0(\tau)$ on the boundary,  Eqs.\ (\ref{eq3110})
and (\ref{eq3112})
define the equation of state $p_0=p_0(\rho_0)$ 
in a parametric form.
  Using (\ref{eq3103}) and (\ref{eq207}) the 
scale $a$ on the RSII brane  can be expressed as 
\begin{equation}
a^2=\frac{\ell^2}{z_{\rm br}^2} a_0^2 Q^2(z_{\rm br}) ,
 \label{eq3303}
\end{equation}
where
\begin{equation}
Q^2(z_{\rm br})=
\left(1-\frac{\mathcal{H}_0^2 z_{\rm br}^2}{4}\right)^2
+ \frac14 \frac{\mu z_{\rm br}^4}{a_0^4},
 \label{eq3221}
\end{equation}
Thus, given the RSII-brane position $z_{\rm br}$ the mapping from the holographic to RSII cosmology is unique.

\begin{figure*}[ht]
\begin{center}
\includegraphics[width=0.45\textwidth ]{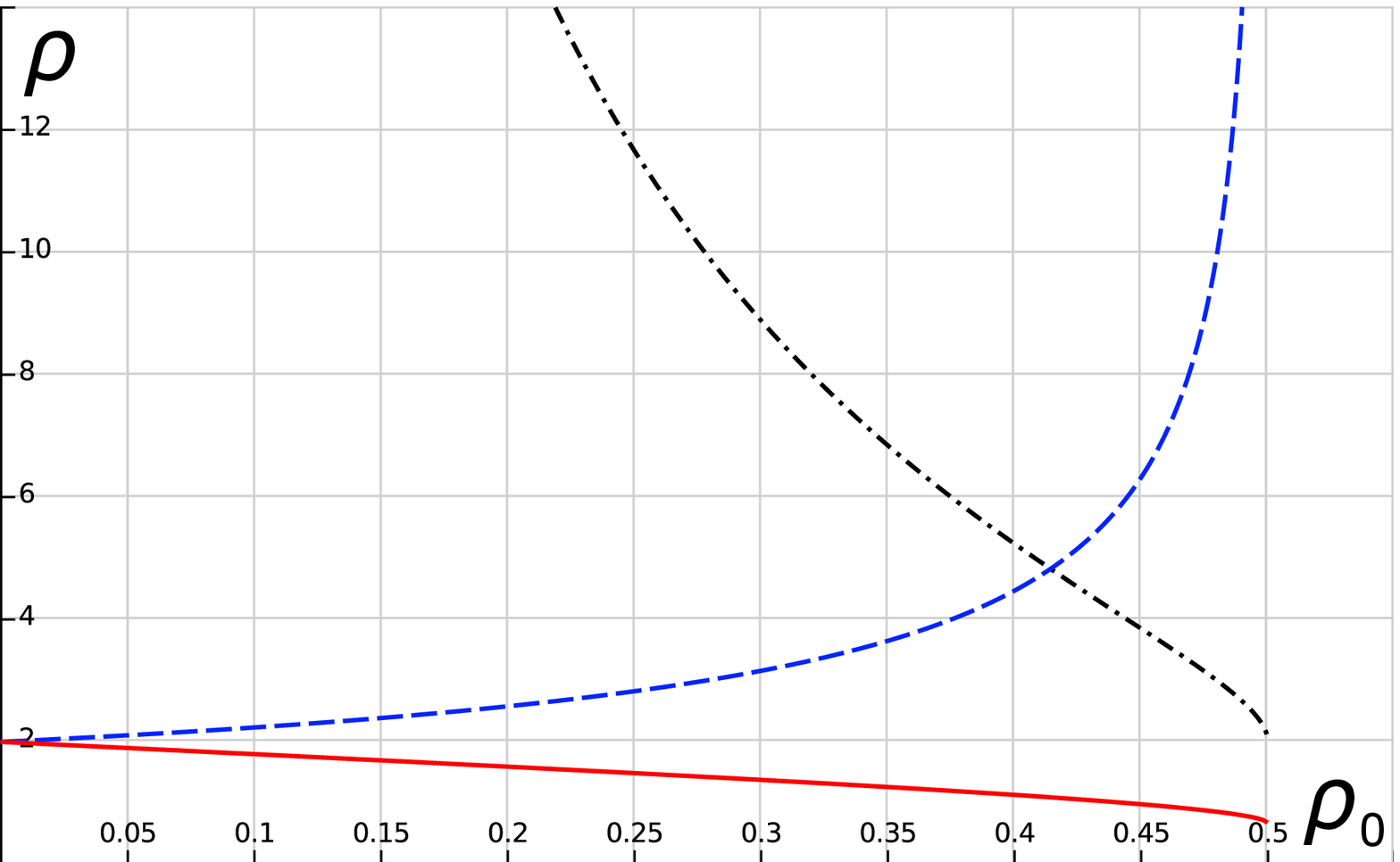}
\hspace{0.02\textwidth}
\includegraphics[width=0.45\textwidth]{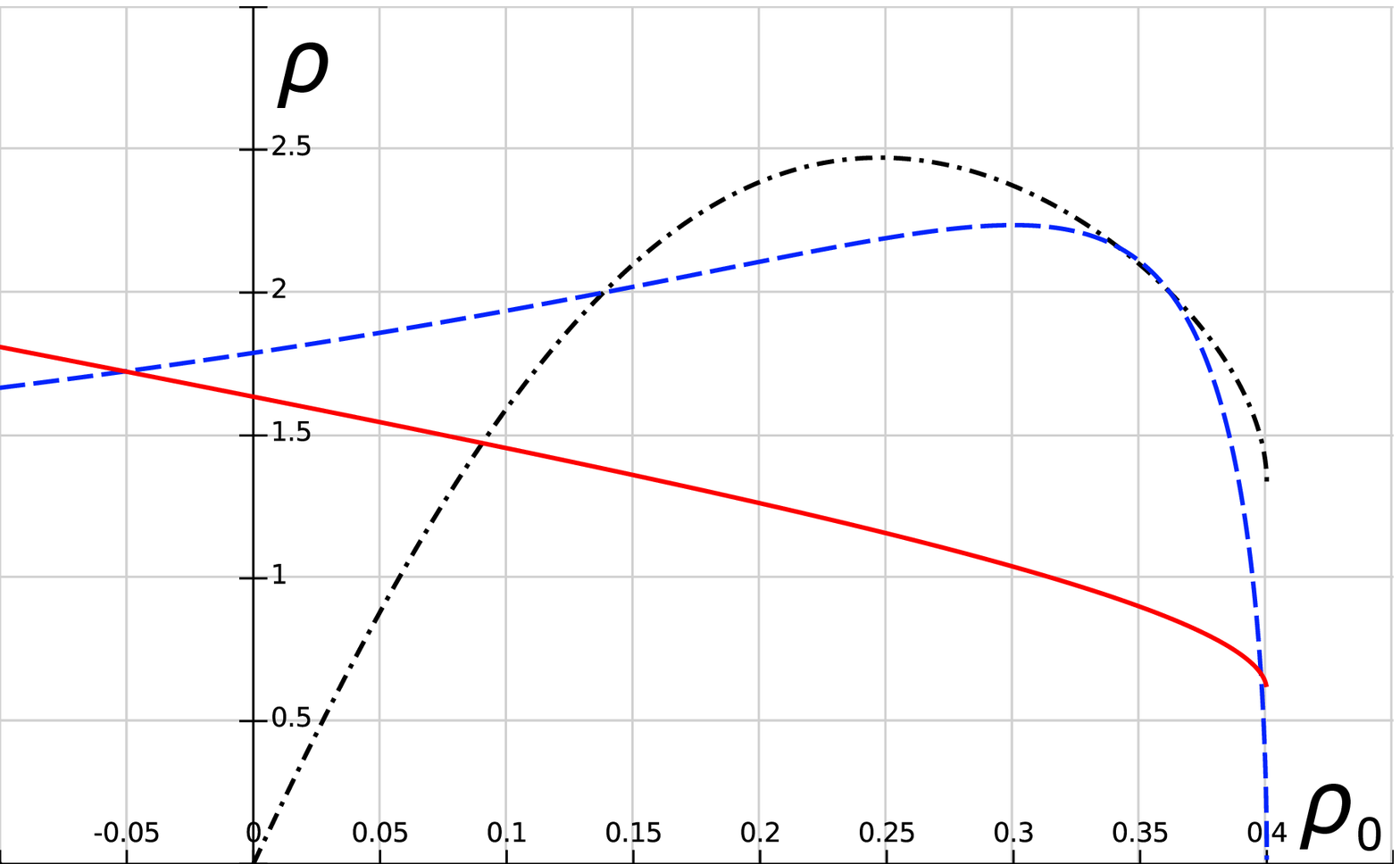}
\\
\vspace{0.02\textwidth}
\includegraphics[width=0.45\textwidth ]{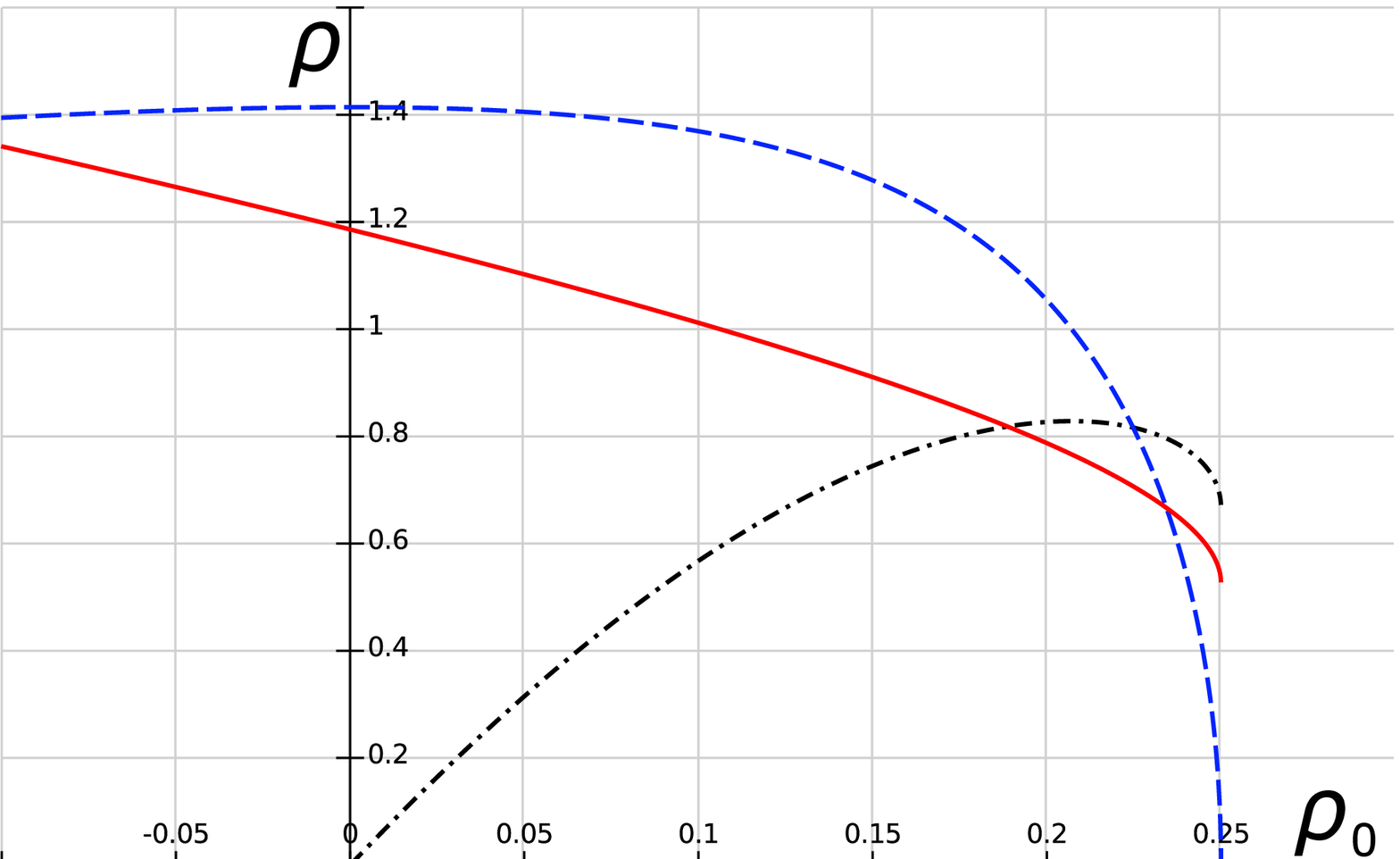}
\hspace{0.02\textwidth}
\includegraphics[width=0.45\textwidth]{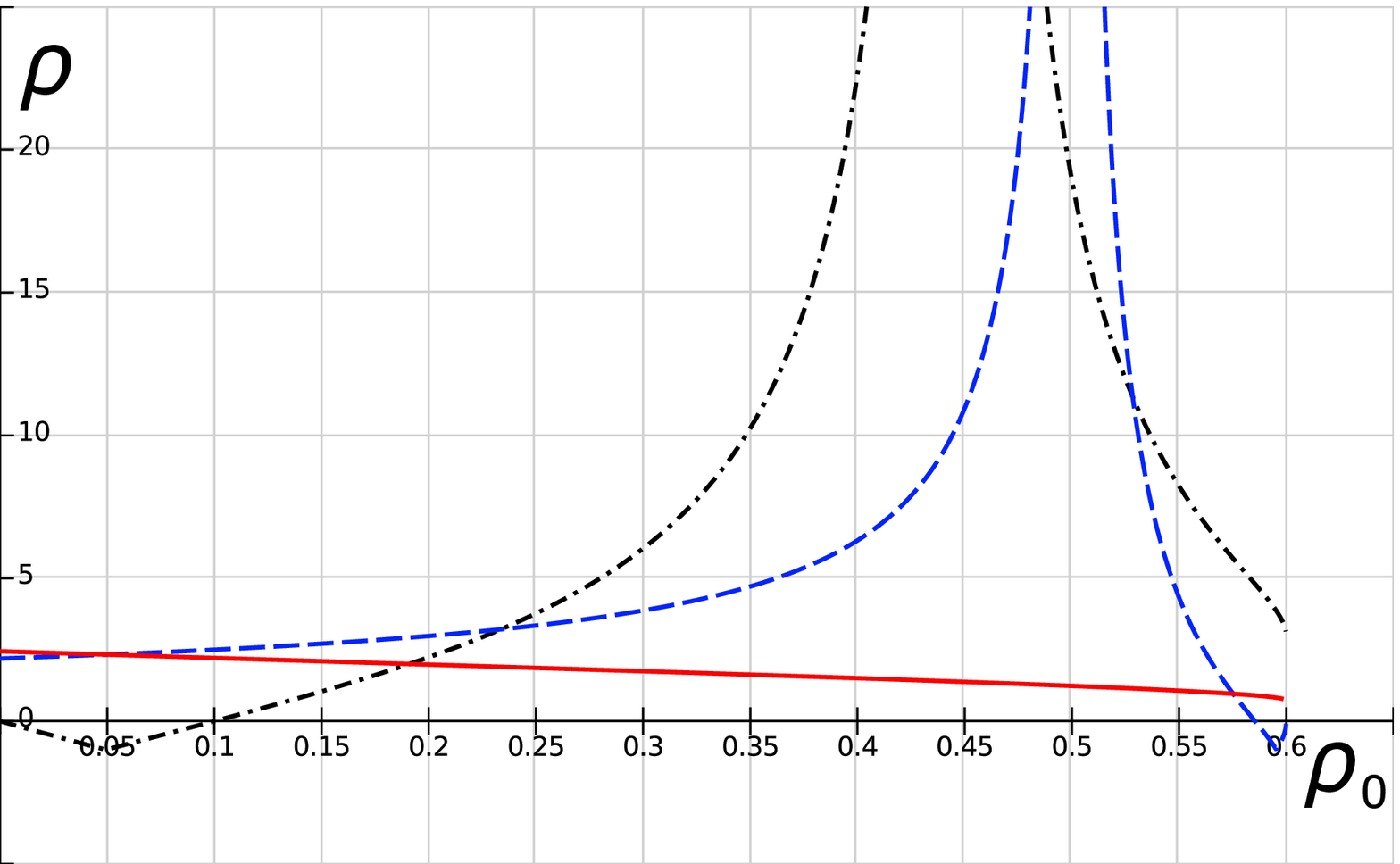}
\caption{Same as in Fig.\ \ref{fig3} for  $\epsilon=+1$.
}
\label{fig4}
\end{center}
\end{figure*}

Knowing $a_0$, we may calculate the effective density of matter on the RSII brane 
assuming the Friedmann equation (\ref{eq032}) holds. Solving (\ref{eq033}) for $\rho$ one finds
\begin{equation}
\rho=\sigma_0\sqrt{1+\mathcal{H}_{\rm br}^2 z_{\rm br}^2
-\frac{\mu z_{\rm br}^4}{\mathcal{A}_{\rm br}^4}} -\sigma .
 \label{eq3314}
\end{equation}
 Then, by making use of (\ref{eq3211}) and (\ref{eq3303})  we obtain the effective density $\rho$ 
on the RSII brane 
expressed in terms of
  $a_0$ and $\mathcal{H}_0$ on the holographic brane: 
 \begin{equation}
\rho=\sigma_0\sqrt{1+\frac{\mathcal{H}_0^2 z_{\rm br}^2}{Q^2(z_{\rm br})}
-\frac{\mu z_{\rm br}^4}{a_0^4Q^4(z_{\rm br})}} -\sigma ,
 \label{eq3316}
\end{equation}
where the function $Q(x)$ is defined by (\ref{eq3221}). 
Using (\ref{eq3115}) and (\ref{eq3211}) we can also express $\rho$ as an explicit function of $a_0$ and $\rho_0$:
\begin{eqnarray}
\frac{\rho}{\sigma_0}&=&\left[1+ 2\frac{z_{\rm br}^2}{\ell^2}\left(
 1+\epsilon \sqrt{1-\frac{2\rho_0}{\sigma_0}-\frac{\mu\ell^4}{a_0^4}} \right)Q^{-2}
 \right.
 \nonumber
 \\
 &&
 \left.
-\frac{\mu z_{\rm br}^4}{a_0^4}Q^{-4}\right]^{1/2} -\frac{\sigma}{\sigma_0} ,
 \label{eq3321}
\end{eqnarray}
where
\begin{eqnarray}
Q^2&=&\left[1- \frac{z_{\rm br}^2}{2\ell^2}\left(
 1+\epsilon \sqrt{1-\frac{2\rho_0}{\sigma_0}-\frac{\mu\ell^4}{a_0^4}} \right)\right]^2
\nonumber
 \\
 && 
+ \left(\frac{z_{\rm br}^2}{2\ell^2}\right)^2\frac{\mu \ell^4}{a_0^4} .
 \label{eq3322}
\end{eqnarray}
Equation (\ref{eq3321}) simplifies considerably when the brane is placed at $z_{\rm br}=\ell$.
In this case we find
\begin{equation}
\frac{\rho}{\sigma_0}=\left|\frac{1+\rho_0/\sigma_0-
 \epsilon \sqrt{1-2\rho_0/\sigma_0-\mu\ell^4/a_0^4}}{1-\rho_0/\sigma_0-
 \epsilon \sqrt{1-2\rho_0/\sigma_0-\mu\ell^4/a_0^4}}\right|
 -\frac{\sigma}{\sigma_0} .
 \label{eq3330}
\end{equation}
Note that the function  $\rho=\rho(\rho_0,a_0)$  is not uniquely defined although 
 the mapping $a_0 \rightarrow a$ is unique.

With no black hole in the bulk, i.e., for $\mu=0$, the density $\rho$ is a function of $\rho_0$ only.
For $\mu\neq 0$ the density $\rho$  depends on both $\rho_0$ and $a_0$.
If the  function $a_0=a_0(\tau)$ is known or the equation of state $p=p(\rho_0)$ is specified
the scale $a_0$ will be an implicit function of $\rho_0$ through the Friedmann equations
(\ref{eq3110}) and (\ref{eq3113}).
However as we have specified neither the equation of state nor the function $a_0=a_0(\tau)$
we  treat $a_0$ as a parameter and show the functional dependence  $\rho=\rho(\rho_0)$
for various values of $\mu \ell^4/a_0^4$ and various $z_{\rm br}^2/\ell^2$ in Figs. \ \ref{fig3} and  \ref{fig4} 
for $\epsilon =-1$ and $+1$, respectively.

It is of interest to analyze the expression (\ref{eq3321}) in the three limiting regimes:  $\left|\rho_0/\sigma_0\right| \ll 1$,  
 $\rho_0/\sigma_0 \rightarrow -\infty$ and $z_{\rm br}\ll \ell$.
  Consider first the regime of large negative $\rho_0$.
Taking the limit $\rho_0/\sigma_0 \rightarrow -\infty$ of (\ref{eq3321}) we find
\begin{equation}
\frac{\rho}{\sigma_0}=\epsilon \frac{4\ell^2}{z_{\rm br}^2}
\sqrt{- \frac{\sigma_0}{2\rho_0}}.
 \label{eq3323}
\end{equation}
This equation is equivalent to Eq.\ (\ref{eq3224}) and 
can be obtained directly from (\ref{eq3224}) by making use of 
the low -density limit of the RSII Friedmann equation (\ref{eq029}) 
and the large $\mathcal{H}_0$ limit of the holographic 
Friedmann equation (\ref{eq3110}).

Next, consider the limit $z_{\rm br}/\ell\ll 1$.
This case is important because, as discussed in Appendix \ref{connection}, in the limit $z_{\rm br}\rightarrow 0$ the RSII brane 
provides an infrared cutoff regularization of the on-shell bulk action.
According to (\ref{eq4007}) in this limit $\mathcal{H}_0|_{\rm max} \rightarrow \infty$
so the necessary condition (\ref{eq4009}) for the $\mathcal{E}(0)=-1$ cannot be met. 
Hence the limit $z_{\rm br}/\ell\ll 1$ is relevant only for the $\mathcal{E}(0)=+1$ branch of the one-sided version.
In this limit $Q(z_{\rm br}) \rightarrow 1$ and the expression (\ref{eq3321})
reduces to  
\begin{equation}
\rho=\sigma_0 -\sigma,
 \label{eq3325}
\end{equation}
so, with the fine-tuning condition $\sigma=\sigma_0$, the effective density on the RSII brane vanishes as 
the brane position approaches the boundary at $z=0$.

\subsubsection{Low -density regime} 
 
 The regime in which
 $\left|\rho_0/\sigma_0\right| \ll 1$ 
is relevant for the one-sided version only since
in this case  $\mathcal{H}_0 \ll 1$ and  
the necessary condition (\ref{eq4009}) for the two-sided version is not met unless  $z_{\rm br}\gg \ell$.  
For $\epsilon=-1$,  $\rho_0\ll \sigma_0$  and  $\mu \ell^4/a_0^4 \ll 1$ we find at linear order in $\mu$
and quadratic order in $\rho_0$
 \begin{eqnarray}
 \frac{\rho}{\sigma_0} &=& 1- \frac{\sigma}{\sigma_0}
 + \frac{z_{\rm br}^2}{\ell^2}\frac{\rho_0}{\sigma_0}
 +\frac12  \frac{z_{\rm br}^2}{\ell^2}\left(\frac{z_{\rm br}^2}{\ell^2}+1\right)\frac{\rho_0^2}{\sigma_0^2}
 \nonumber
 \\
 &&
 -\frac12  \frac{z_{\rm br}^2}{\ell^2}\left(\frac{z_{\rm br}^2}{\ell^2}-1\right)\frac{\mu \ell^4}{a_0^4}  + \dots .
 \label{eq3317}
\end{eqnarray}
Hence, at linear order the effective energy density on the RSII brane 
equals the energy density on the holographic brane multiplied by a constant plus  the cosmological constant term
which can be eliminated by adopting the 
RSII fine-tuning condition  $\sigma=\sigma_0$.

The effective pressure on the RSII brane can be easily derived by making use of the energy conservation equations 
(\ref{3201}) on the RSII brane and (\ref{3200}) on the holographic brane.
At linear order one finds
\begin{equation}
p=-(\sigma_0-\sigma) +\frac{z_{\rm br}^2}{\ell^2} p_0+\dots .
 \label{eq3320}
\end{equation}
Hence, at linear order the effective fluid on the RSII brane 
 satisfies the same equation of state as the fluid on the holographic brane.
The cosmological constant term 
will vanish on both branes if the RSII fine-tuning condition is imposed
whereas the dark radiation contribution will be the same on the two branes  
only if $z_{\rm br}=\ell$.
We recover the standard cosmology  on both branes by choosing $\ell$ 
such that  $\sigma_0$ is sufficiently large to suppress the quadratic and higher terms
in (\ref{eq3317}).

For $\epsilon=+1$ and 
$z_{\rm br}^2/\ell^2=1$, $\rho$ diverges in the limit $\rho_0\rightarrow 0$. 
For an arbitrary $z_{\rm br}^2/\ell^2$ we find at linear order
  \begin{eqnarray}
 \frac{\rho}{\sigma_0}&=&\frac{z_{\rm br}^2/\ell^2 +1}{z_{\rm br}^2/\ell^2 -1}-\frac{\sigma}{\sigma_0}
 +\frac{z_{\rm br}^2/\ell^2}{(z_{\rm br}^2/\ell^2 -1)^2}
  \frac{\rho_0}{\sigma_0}
  \nonumber \\
  &&
- \frac{z_{\rm br}^2/\ell^2}{2(z_{\rm br}^2/\ell^2 -1)^3} \frac{\mu \ell^4}{a_0^4} + \dots .
 \label{eq3324}
\end{eqnarray}
In this case the effective energy density on the RSII brane at linear order 
differs from the energy density on the holographic brane by a multiplicative constant.
Besides, for $\sigma=\sigma_0$ the effective cosmological constant does not vanish and is equal to
\begin{equation}
\Lambda_{\rm br}=\frac{6}{\ell^2}\frac{z_{\rm br}^2/\ell^2 +1}{z_{\rm br}^2/\ell^2 -1}-\frac{6}{\ell^2} .
 \label{eq3312}
\end{equation}
This scenario offers a few interesting possibilities.
Suppose the energy density  $\rho_0$ on the holographic brane describes matter with the equation of state
satisfying $3p_0+\rho_0>0$, as for,  e.g., cold dark matter.
According to (\ref{eq3324}) and (\ref{eq3312}) we  have 
an asymptotically de Sitter universe on the RSII brane.
The location of the brane is crucial.
For $z_{\rm br}$ of the order of $\ell$  (excluding $z_{\rm br}=\ell$)
we could have the  standard $\Lambda$CDM cosmology on the RSII brane 
if we included a  
cosmological constant term in $\rho_0$ and fine tuned it to 
cancel  $\Lambda_{\rm br}$ up to a small phenomenologically acceptable contribution.
In principle, this could work even without the RSII fine-tuning condition $\sigma=\sigma_0$.
For small $\ell/  z_{\rm br}$
if we impose the RSII fine-tuning condition,
 both the constant and linear terms will be suppressed by a factor $\ell^2/  z_{\rm br}^2$.
So  we can choose  
the ratio $\ell/  z_{\rm br}$ such that
the effective cosmological constant  $\Lambda_{\rm br}$   fits the observed value today
\begin{equation}
\Lambda= 3\Omega_\Lambda H_0^2 ,
 \label{eq3313}
\end{equation}
where $H_0$ is today's Hubble constant  of the order of $2.5\times 10^{-40} {\rm GeV}$.
Expanding (\ref{eq3312}) for small $\ell/z_{\rm br}$ and equating $\Lambda_{\rm br}$ with $\Lambda$
we find  
\begin{equation}
\frac{\ell^2}{z_{\rm br}^2}=\frac{\sqrt{\Omega_\Lambda}}{2}H_0\ell\lesssim 10^{-28} ,
 \label{eq3315}
\end{equation}
where  the numerical estimate of the right-hand side is obtained 
for $\Omega_\Lambda\simeq 0.7$ and 
the Newtonian potential constraint (\ref{eq0021}) at small distances 
with   $\ell \lesssim 10^{12} {\rm GeV}^{-1}$.

\subsection{RSII cosmological scenario}
\label{rs2}

In the RSII  scenario the primary braneworld is the RSII brane at $z=z_{\rm br}$ 
with cosmology  determined by Eqs.\
(\ref{eq029}) and (\ref{eq3223}).
Observers at the boundary brane at $z=0$ experience the emergent cosmology which is a reflection
of the RSII cosmology.
We would like to express the cosmological scale and effective energy density on the holographic brane at $z=0$ in terms of
cosmological scale and energy density on the RSII brane.
For simplicity, in the following we assume the RSII fine-tuning condition $\sigma=\sigma_0$.
If the density $\rho$ and pressure $p$ on the RSII brane are known, the
cosmological scale $a$ may be derived by integrating (\ref{eq029}) and (\ref{eq3223}).
On the other hand, given $a(\tau)$ on the RSII brane,  Eqs.\ 
(\ref{eq029}) and (\ref{eq3223})
define the equation of state $p=p(\rho)$ 
in a parametric form.
From  (\ref{eq1101}) we find the 
scale $a_0$ on the holographic brane   expressed in terms of $a$
\begin{equation}
a_0^2 = \frac{a^2}{2}\frac{z_{\rm br}^2}{\ell^2}
\left(
1 +\frac{\mathcal{H}_{\rm RSII}^2 \ell^2}{2} +\mathcal{E}_0
\sqrt{1+\mathcal{H}_{\rm RSII}^2 \ell^2
-\frac{\mu \ell^4}{a^4}} 
\right),
 \label{eq3214}
\end{equation}
where, as before,  $\mathcal{E}_0\equiv \mathcal{E}(0)=-1$ for the two-sided and $\mathcal{E}_0=+1$ or $-1$ for the one-sided version
of the RSII model. Thus the mapping $a\rightarrow a_0$ is unique only for the two-sided model.
 
Knowing $a$, we may calculate the effective density of matter on the holographic brane 
assuming the Friedmann equation (\ref{eq3110}) holds. 
As before, this can be done for an arbitrary $z_{\rm br}$.

Using (\ref{eq032}), (\ref{eq3209}), and (\ref{eq3214}) we can express the Hubble rate $\mathcal{H}_0$ 
and the scale $a_0$ in terms
of $\rho$ and $a$:
\begin{equation}
\mathcal{H}_0^2 = \frac{4}{z_{\rm br}^2}
\frac{(\rho/\sigma_0+1)^2-1+\mu \ell^4/a^4}{(\rho/\sigma_0+1+\mathcal{E}_0)^2+\mu \ell^4/a^4},
 \label{eq3215}
\end{equation}
 \begin{equation}
a_0^2= \frac{a^2}{4}\frac{\ell^2}{z_{\rm br}^2}\left[\left( \frac{\rho}{\sigma_0}+1+\mathcal{E}_0  \right)^2+
\frac{\mu \ell^4}{a^4}  \right].
 \label{eq3216}
\end{equation}
Next, using (\ref{eq3115}) to replace $\mathcal{H}_0^2$ in (\ref{eq3215}), 
substituting the expression (\ref{eq3216}) for $a_0$, and solving for $\rho_0$ we find
\begin{widetext}
\begin{equation}
\frac{\rho_0}{\sigma_0}= 4\frac{\ell^2}{z_{\rm br}^2}
\frac{(\rho/\sigma_0+1)^2-1+(\ell/z_{\rm br})^2 (\rho/\sigma_0+1)(\mathcal{E}_0-\rho/\sigma_0-1)
+(1-\ell^2/z_{\rm br}^2)\mu \ell^4/a^4
}{(\rho/\sigma_0+1+\mathcal{E}_0)^2+\mu \ell^4/a^4} .
 \label{eq3217}
\end{equation}
\end{widetext}
To simplify the analysis
consider $z_{\rm br}=\ell$. 
 For the two-sided RSII model along with the $\mathcal{E}_0=-1$ branch of the one-sided model we have
\begin{equation}
\frac{\rho_0}{\sigma_0}= -4 \frac{\rho/\sigma_0+2}{(\rho/\sigma_0^2)^2+ \mu\ell^4/a^4} .
 \label{eq3218}
\end{equation}
Thus, the two-sided model with positive energy density $\rho$ and positive $\mu$ maps
into a holographic cosmology with negative effective energy density $\rho_0$. For $\mu=0$ the density
$\rho_0$  diverges for small $\rho$ as 
$1/\rho$. 
The one-sided model maps into two branches: the $\mathcal{E}_0=-1$ branch  identical with the two-sided map
and  the  $\mathcal{E}_0=+1$ branch in which case we find
\begin{equation}
\frac{\rho_0}{\sigma_0}=  \frac{4\rho/\sigma_0}{(\rho/\sigma_0^2+2)^2+ \mu\ell^4/a^4} ,
 \label{eq3219}
\end{equation}
yielding smooth positive $\rho_0$.
Note that the inverse function  $\rho=\rho(\rho_0)$ of (\ref{eq3217}) for $\mu=0$ and $z_{\rm br}=\ell$ 
coincides with the function  defined by (\ref{eq3330}) for $\mu=0$ if  we set $\mathcal{E}_0=+1$ for $\rho_0>0$
and $\mathcal{E}_0=-1$ for 
 $\rho_0<0$.

\section{Summary and conclusions}
\label{conclusion}

We have explicitly constructed the holographic mapping between two cosmological braneworld scenarios: 
holographic  and RSII braneworld.
In the holographic scenario the primary braneworld is at the boundary of AdS$_5$
with emergent cosmology at the RSII braneworld.
In the RSII scenario the primary braneworld is located at an arbitrary nonzero $z=z_{\rm br}$
and the cosmology at the $z=0$ boundary is emergent.
In both scenarios we have established a holographic map between these two braneworld cosmologies.

We have assumed the presence of matter on the primary braneworld only.
The emergent cosmology is governed by the Friedman equations with effective
energy density and pressure.
We have obtained functional relations between cosmological scales $a_0$ and $a$,
Hubble rates $H_0$ and $H$ and effective energy densities $\rho_0$ and $\rho$ in the two scenarios.
We have analyzed two versions of the RSII models:
the so called ``one-sided'' and ``two-sided'' version.
We have demonstrated that the  map between the cosmological scales is unique for the two-sided RSII model
whereas in the one-sided model the mapping from the holographic to the RSII cosmology
is a two-valued function. 

In particular, we have studied the low density regime, i.e., the regime in which 
$\rho\simeq \rho_0 \ll 1/(G_{\rm N}\ell^2)$.
 The low density regime can be made simultaneous only in the one-sided RSII model
since the necessary condition (\ref{eq4008}) for the two-sided version 
is not met if both $\rho_0$ and $\rho$ are small. 
The low-density regime on the two-sided RSII brane corresponds to
the high negative energy density limit on the holographic brane.

The analysis presented here is open to speculations.
For example, it is conceivable that our Universe is a one-sided RSII braneworld
the cosmology of which is emergent from the primary holographic cosmology.
If $\rho_0$ on the holographic brane describes matter with the equation of state
satisfying $3p_0+\rho_0>0$, as for,  e.g., cold dark matter,  in the one-sided model, we will,
according to (\ref{eq3324}) and (\ref{eq3312}), have 
an asymptotically de Sitter universe on the RSII brane. 
With the AdS curvature $\ell$ satisfying the Newtonian potential constraint 
if we choose an appropriate brane location so that $\Lambda_{\rm br}$ fits the observed value today, 
we could produce the standard $\Lambda$CDM cosmology on the RSII brane.
Unfortunately in this scenario we have to push the brane as far as $10^{28}\ell$ 
away from the boundary which seems rather unnatural. 
Another way to  recover the standard cosmology is to involve a 
negative holographic brane tension  in addition to $\rho_0$ and fine tune it to 
cancel  $\Lambda_{\rm br}$ up to a small phenomenologically acceptable contribution.
 
\subsection*{Acknowledgments}
I am indebted to Nikolaos Tetradis, Elias Kiritsis, and Silvije Domazet for useful comments
and to Dijana\ Toli\'c for technical assistance.
This work has been supported by the Croatian 
Science Foundation under Project No.\ IP-2014-09-9582 and
supported in part by the ICTP-SEENET-MTP Grant
No.\ PRJ-09 Strings and Cosmology in the frame of the
SEENET-MTP Network.

\appendix

\section{Second Randall-Sundrum model}
\label{rsII}

Our curvature conventions are as follows:
$R^{a}{}_{bcd} = \partial_c \Gamma_{db}^a - 
\partial_d \Gamma_{cb}^a + \Gamma_{db}^e \Gamma_{ce}^a  - \Gamma_{cb}^e \Gamma_{de}^a$ 
and $R_{ab} = R^s{}_{asb}$, 
so that Einstein's equations are $R_{ab} - \frac{1}{2}R G_{ab} = +8\pi G T_{ab}.$ 
The dynamics of a 3-brane in a 4+1-dimensional bulk is described by 
the total action as the sum of the bulk and  brane actions  
\begin{equation}
 S=S_{\rm bulk}+S_{\rm br} .
 \label{eq0002}
\end{equation}
The bulk action is given by
\begin{equation} 
S_{\rm bulk} =\frac{1}{8\pi G_5} \int d^5x \sqrt{G} 
\left[-\frac{R^{(5)} }{2} -\Lambda _5 \right] 
+S_{\rm GH}[h],
\label{eq001} 
\end{equation}
where $\Lambda_5$ is the bulk cosmological constant related to the 
AdS curvature radius as $\Lambda_5=-6/\ell^2$.
The Gibbons-Hawking boundary term is given by an integral over the brane hypersurface $\Sigma$:
\begin{equation} 
S_{\rm GH}[h] =\frac{1}{8\pi G_5}\int_\Sigma d^{4}x\sqrt{-h} K[h] .
\label{eq003}
\end{equation} 
The quantity $K$ is the trace of the extrinsic curvature tensor $K_{ab}$
defined as 
\begin{equation}
K_{ab}=h_{a}^{c}h_{b}^{d} n_{d;c} ,
\label{eq109}
\end{equation}
where $n^a$ is a unit vector normal to the brane
pointing toward increasing $z$,
$h_{ab}$ is 
 the induced metric
\begin{equation}
 h_{ab}=G_{ab}+n_a n_b ,
 \label{eq1003}
\end{equation}
and $h\equiv \det h_{ab}$ is its determinant.
Observers reside on the positive tension brane with action
\begin{equation} 
S_{\rm br}[h] =\int d^{4}x\sqrt{-h} (-\sigma + \mathcal{L}^{\rm matt}[h]),
\label{eq1005}
\end{equation} 
  where they see the metric $h_{\mu \nu }$.

The basic equations are
  the bulk field equations outside the brane
\begin{equation}
R^{(5)}_{ab}-\frac12 R^{(5)}G_{ab}= \Lambda_5 G_{ab}
 \label{eq0003}
\end{equation}
and junction conditions \cite{israel}
\begin{equation}
\left[ \left[ K^{\mu}_{\nu} - \delta^{\mu}_{\nu}  K_{\alpha}^{\alpha} \right] \right]
= 8 \pi G_5  (\sigma\delta^{\mu}_{\nu}+ T^{\mu}_{\nu}) ,
\label{eq009}
\end{equation}
where the energy-momentum tensor $T^{\mu}_{\nu}={\rm diag} (\rho,-p,-p,-p)$ describes matter on the brane  
and $[[f]]$ denotes the discontinuity of a function $f(z)$ across the brane,
i.e.,
\begin{equation}
\left[ \left[ f (z) \right] \right] =
\lim_{\epsilon \rightarrow 0}  \,
\left( f (z_{\rm br} + \epsilon) - f (z_{\rm br}-\epsilon) \right) .
\end{equation}

To derive the RSII model solution it is convenient to 
use Gaussian normal coordinates $x_a=(x_\mu,y)$ with the fifth  
coordinate $y$ related to the Fefferman-Graham coordinate $z$ by $z=\ell e^{y/\ell}$.
Then, in the two-sided version with the  $Z_2$ symmetry 
$y-y_{\rm br}\leftrightarrow y_{\rm br}-y$ one identifies the region
$-\infty<y\leq y_{\rm br}$ with $ y_{\rm br}\leq y <\infty$.
We start with a simple ansatz for the line element 
\begin{equation}  
ds_{(5)}^{2} =\psi^2(y)g_{\mu \nu }(x) dx^{\mu } dx^{\nu } -dy^{2} ,
\label{eq002}
\end{equation} 
where the warp factor $\psi^2$ is a function of $y$. We assume 
that $\psi^2\rightarrow 0$ as
as $y\rightarrow \infty$ and 
\begin{equation}
\psi^2(y_{\rm br})=1.
\label{eq005}
\end{equation}
Then, the total action (\ref{eq003}) may be brought to the form \cite{kim,bilic}
\begin{eqnarray} 
S[g] &=&\frac{1}{8\pi G_5} \int d^4x \sqrt{-g}\int dy 
\left[-\frac{R}{2}\psi^2 -4(\psi^3\psi')'
\right.
\nonumber\\
&&
\left.
+6 \psi^2 (\psi')^2-\Lambda ^{(5)}\psi^4 \right]
+S_{\rm GH}[g]+S_{\rm br}[g] ,
\label{eq008} 
\end{eqnarray}
where $R$ is the four-dimensional Ricci scalar associated with the metric $g_{\mu\nu}$ and
the prime $'$ denotes a derivative with respect to $y$.
The extrinsic curvature is easily calculated using the definition (\ref{eq109}) and 
the unit normal vector $n^\mu=(0,0,0,0,1)$. We find the nonvanishing components
\begin{equation} 
K_{\mu\nu}=n_{\mu;\nu}=
-\Gamma^a_{\mu\nu}n_a=\psi \psi' g_{\mu\nu} .
\label{eq004} 
\end{equation}
The fifth coordinate in (\ref{eq008} )  may be integrated out 
if $\psi\rightarrow 0$ sufficiently fast as 
we approach $y=\infty$.

The functional form of $\psi$ is found by solving the Einstein equations (\ref{eq0003})
outside the brane. Using the components of the Ricci tensor 
\begin{equation}
R^{(5)}_{55}=-4\frac{\psi''}{\psi}, \quad  R^{(5)}_{5\mu}=0,
 \label{eq0004}
\end{equation}
\begin{equation}
R^{(5)}_{\mu\nu}=R_{\mu\nu}+\left(3{\psi'}^2+\psi\psi''
\right) g_{\mu\nu} ,
 \label{eq0005}
\end{equation}
and the Ricci scalar
\begin{equation}
R^{(5)}=\frac{R}{\psi^2}+12\frac{{\psi'}^2}{\psi^2}+8\frac{\psi''}{\psi} ,
 \label{eq0006}
\end{equation}
we find the 55 and $\mu\nu$  components of the Einstein equations,
respectively, as 
\begin{equation}
6\frac{{\psi'}^2}{\psi^2}+\Lambda^{(5)}+\frac{R}{2\psi^2}=0
 \label{eq0007}
\end{equation}
and 
\begin{equation}
R_{\mu\nu}- \frac12 R g_{\mu\nu}=\left(3{\psi'}^2+3\psi\psi''+\Lambda^{(5)}\psi^2
\right) g_{\mu\nu} .
 \label{eq0008}
\end{equation}
Combining Eq.\ (\ref{eq0007}) with the  contracted Eq.\ (\ref{eq0008}) we find
\begin{equation}
\psi=e^{- (y-y_{\rm br})/\ell} ,
 \label{eq0009}
\end{equation}
as the unique solution to Eqs. (\ref{eq0003}) which satisfies the condition (\ref{eq005}) and vanishes at $y=\infty$, 
where   $\ell=\sqrt{-6/\Lambda^{(5)}}$.
With this solution, the metric (\ref{eq002}) is AdS$_5$ in normal coordinates 
because the constant factor $e^{y_{\rm br}/\ell}$ may be removed by a coordinate translation 
$y \rightarrow \tilde{y}=y-y_{\rm br}$.
Equation (\ref{eq0008}) then reduces to the four-dimensional Einstein equation in empty space
\begin{equation}
R_{\mu\nu}- \frac12 R g_{\mu\nu}=0.
 \label{eq0010}
\end{equation}
This equation should follow from the variation of the action (\ref{eq008}) with ${\mathcal L}_{\rm matt}=0$
after integrating out the fifth coordinate.
For this to happen it is necessary that the last three terms in square brackets
are canceled by the boundary term and the brane action without matter.
Using (\ref{eq004}) one finds that  the integral of the second term in square brackets is precisely canceled 
by the Gibbons-Hawking term. Then, the cancellations of the remaining terms will take place if
\begin{equation} 
\frac{\gamma}{8\pi G_5}\int^\infty_{y_{\rm br}} dy 
\left[6 \psi^2 (\psi')^2-\Lambda ^{(5)}\psi^4 \right] =\sigma ,
\label{eq0011} 
\end{equation}
where
\begin{equation}
\gamma=\left\{ \begin{array}{ll}
1, & \mbox{one-sided RSII},\\
2, & \mbox{two-sided RSII}.\end{array} \right.\
\label{eq2003}
\end{equation}
This equation yields
the  RSII fine-tuning condition
\begin{equation} 
\sigma=\sigma_0\equiv \frac{3\gamma}{8\pi G_5\ell}.
\label{eq0012} 
\end{equation}
In this way, after integrating out the fifth dimension, the total effective four-dimensional action 
assumes the form of the standard
Einstein-Hilbert action without cosmological constant
\begin{equation} 
S=\frac{1}{8\pi G_{\rm N}}\int d^4x \sqrt{-g}\left(-\frac{R}{2}\right) ,
\label{eq0013} 
\end{equation}
where $G_{\rm N}$ is the Newton constant defined by
\begin{equation} 
\frac{1}{G_{\rm N}}= \frac{\gamma}{G_5}\int^\infty_{y_{\rm br}}  dy \psi^2=\frac{\gamma\ell}{2G_5}.
\label{eq0014} 
\end{equation}
Then, the constant $\sigma_0$  in (\ref{eq0012}) is given by
\begin{equation} 
\sigma_0= \frac{3}{4\pi G_{\rm N}\ell^2} ,
\label{eq2005} 
\end{equation}
so the RSII fine-tuning condition does not depend on the sidedness $\gamma$ if $\sigma_0$ is
expressed in terms of the four-dimensional Newton constant.

It is worth noting that the fine-tuning condition (\ref{eq0012}) and (\ref{eq0014}) follows directly
from the junction conditions   (\ref{eq009}) and the metric (\ref{eq002}) with (\ref{eq0009}).
For a static brane at $y=y_{\rm br}$ we find 
\begin{equation} 
K_{\mu\nu}|_{y=y_{\rm br}+\epsilon}=-\frac{1}{\ell}g_{\mu\nu}.
\label{eq1015} 
\end{equation}
For the one-sided version we can set 
\begin{equation} 
K_{\mu\nu}|_{y=y_{\rm br}-\epsilon}=0,
\label{eq2015} 
\end{equation}
whereas  the  two-sided version 
or the $Z_2$ symmetry implies
\begin{equation} 
K_{\mu\nu}|_{y=y_{\rm br}-\epsilon}=-K_{\mu\nu}|_{y=y_{\rm br}+\epsilon} .
\label{eq0015} 
\end{equation}
Then, 
from  (\ref{eq009}) we find
\begin{equation} 
\frac{3\gamma}{\ell}\delta^\mu_\nu = 8\pi G_5 \sigma \delta^\mu_\nu ,
\label{eq0016} 
\end{equation}
yielding (\ref{eq0012}). 

Equation (\ref{eq0010}) admits any Ricci flat metric as its solution. The trivial solution  $g_{\mu\nu}=\eta_{\mu\nu}$ 
gives the original RSII model \cite{randall2}
with an empty Minkowski brane located at an arbitrary $y=y_{\rm br}$ in the AdS/$Z_2$ bulk.
More general solutions with a black hole on the brane are first considered in Ref. \cite{chamblin2} 
and discussed in more detail in Ref.\ \cite{emparan}.

The RSII model can be extended to include  a brane with spherical
or hyperbolic geometry embedded in the AdS-Schwarzschild geometry
\cite{gomez,birmingham}. 
In this case it is convenient to represent the bulk metric 
in  Schwarzschild coordinates (\ref{eq3202}).
It is worth mentioning that
the solution (\ref{eq3202}) is closely related 
to the D3-brane solution of ten-dimensional supergravity 
corresponding to a stack of $N_{\rm D}$ coincident D3-branes. If we identify the AdS curvature radius  with
$\ell=\ell_s (4\pi g_sN_{\rm D})^{1/4}$, where
$g_s$ is 
the string coupling constant, and 
 $\ell_s=\sqrt{\alpha'}$ is the fundamental string length,
a  near-horizon nonextremal D3-brane metric is given by  \cite{horowitz}
\begin{equation}
ds_{(10)}^{2} = ds_{\rm ASch}^2
-\ell^2 d\Omega_5^2 .
\label{eq3017}
\end{equation}

The coordinates $r$ and $z$ are related by
\begin{equation}
\frac{r^2}{\ell^2}=  \frac{\ell^2}{z^2}- \frac{\kappa}{2} +\frac{\kappa^2+4\mu}{16}\frac{z^2}{\ell^2}.
\label{eq3203}
\end{equation}
The brane is placed at $z_{\rm br} <z_{\rm h}$ corresponding to the 
$r_{\rm br} >r_{\rm h}$. The location of the horizon   $r_{\rm h}$ is the positive solution to
the equation $f(r)=0$ yielding
\begin{equation}
r_{\rm h}^2=\frac{\ell^2}{2}(\sqrt{\kappa^2+4\mu}-\kappa), \quad z_{\rm h}^2=\frac{4\ell^2}{\sqrt{\kappa^2+4\mu}}.
\label{eq0017}
\end{equation}
The fifth coordinate is cut at the horizon \cite{gomez} so the bulk in the one-sided version 
is the section of spacetime
defined by $z_{\rm br}\leq z<  z_{\rm h}$.
In the two-sided version 
 one identifies the region $z_{\rm br}\leq z<  z_{\rm h}$ with
$z_{\rm br}^2/z_{\rm h}<  z\leq z_{\rm br}$ with a fixed point at $z= z_{\rm br}$.
Note that the RSII braneworld may be arbitrarily close to the AdS boundary
since $z_{\rm br}$ can be chosen  arbitrarily small but not zero.

The junction conditions for the brane placed at $r_{\rm br}$ 
yield two independent equations
\begin{equation}
\frac{f^{1/2}(r_{\rm br})}{r_{\rm br}}=\frac{1}{\ell}\left(1+\kappa\frac{\ell^2}{r_{\rm br}^2} 
-\mu \frac{\ell^4}{r_{\rm br}^4}\right)^{1/2}=\frac{8\pi G_5}{3\gamma}\sigma,
\label{eq006}
\end{equation}
\begin{eqnarray}
\left. \frac{1}{2f^{1/2}(r_{\rm br})}\frac{df}{dr}\right|_{r=r_{\rm br}} &=&\frac{1}{f^{1/2}(r_{\rm br})}
\left(\frac{r_{\rm br}}{\ell^2} +\mu \frac{\ell^2}{r_{\rm br}^3}\right)
\nonumber \\
&=&\frac{8\pi G_5}{3\gamma}\sigma.
\label{eq007}
\end{eqnarray}
Solving these equations for $r_{\rm br}^2$ and $\sigma$ we obtain 
\begin{equation}
r_{\rm br}^2=\frac{2\mu\ell^2}{\kappa},
 \label{eq010}
\end{equation}
\begin{equation}
\sigma=\sigma_0 \left( 1+\frac{\kappa^2}{4\mu}\right)^{1/2}.
\label{eq011}
\end{equation}
Clearly, for $\kappa=0$ we must have $\mu=0$, in which case we recover  
the standard RSII with flat brane at an arbitrary $r_{\rm br}=\ell^2/z_{\rm br}$, and
Eq.\ (\ref{eq011}) reduces to the fine-tuning condition (\ref{eq0012}).
In contrast, in the case of $\kappa^2=1$ the  brane location  is fixed by (\ref{eq010})
with the requirement that $\mu$ is positive or negative for positive or negative $\kappa$, respectively.

Next, we give a simple derivation of the RSII braneworld cosmology
following  Soda \cite{soda}.
We start from the bulk line element 
in  Schwarzschild coordinates (\ref{eq3202})
and allow the brane  
to move in the bulk along the fifth dimension $r$.
In other words, the brane hypersurface $\Sigma$ 
is time dependent and may be
defined by 
\begin{equation}
r-a(t)=0 ,
\label{eq027}
\end{equation}
where $a=a(t)$ is an arbitrary function.
The normal to $\Sigma$ is then given by
\begin{equation}
n_\mu \propto\partial_\mu (r-a(t))=(-\partial_t a,0,0,0,1)
\label{eq013}
\end{equation}
and, using the normalization $g^{\mu\nu}n_\mu n_\nu=-1$, one finds the nonvanishing components
\begin{equation}
n_t =-\frac{f^{1/2}\partial_t a}{(f^2-(\partial_t a)^2)^{1/2}}, 
\label{eq014}
\end{equation}
\begin{equation}
n_r =\frac{f^{1/2}}{(f^2-(\partial_t a)^2)^{1/2}} ,
\label{eq0141}
\end{equation}
where the function $f$ is given by (\ref{eq3225}) with $r$ replaced by $a$, i.e.,
\begin{equation}
f(a)=\frac{a^2}{\ell^2}+\kappa -\mu \frac{\ell^2}{a^2} .
\label{eq101}
\end{equation}
Using this, from  (\ref{eq014}), (\ref{eq0141}), and (\ref{eq1003}) we find the induced line element on the brane 
\begin{equation}
ds_{\rm ind}^2=n^2(t)dt^2 -a(t)^2 d\Omega_\kappa^2 ,
\label{eq028}
\end{equation}
where 
\begin{equation}
n^2 =f-\frac{(\partial_t a)^2}{f} .
\label{eq016}
\end{equation}
Assuming that either the relation (\ref{eq2015}) or (\ref{eq0015}) holds for the dynamical brane,
the junction conditions (\ref{eq009}) may be written in the form
\begin{equation}
 \left. K_{\mu\nu}\right|_{r=a-\epsilon}=\frac{8 \pi G_5}{3\gamma} \left[3T_{\mu\nu}- (\sigma +T) g_{\mu\nu}
 \right]  .
\label{eq018}
\end{equation}
Then, the $\chi\chi$ component gives
\begin{equation}
 \frac{f^{3/2}}{(f^2-(\partial_t a)^2)^{1/2}}=\frac{8 \pi G_5}{3\gamma}(\sigma+\rho)a.
\label{eq019}
\end{equation}
It turns out that the $tt$ component gives the time derivative of the above equation
and hence  imposes no additional constraint.
Using (\ref{eq016}), Eq.\ (\ref{eq019}) may be cast into the form
\begin{equation}
\frac{(\partial_t a)^2}{n^2 a^2} +\frac{f}{a^2}=\frac{1}{\ell^2 \sigma_0^2}(\sigma+\rho)^2 .
\label{eq020}
\end{equation}
The first term on the left-hand side of (\ref{eq020}) is the square of 
the Hubble expansion rate for the metric (\ref{eq028}) on the brane
\begin{equation}
 H_{\rm RSII}^2=\frac{(\partial_t a)^2}{n^2 a^2}.
\label{eq021}
\end{equation}
Substituting for $f$ the expression (\ref{eq101}) into (\ref{eq020}),    
we obtain the effective Friedmann equation
\begin{equation}
 H_{\rm RSII}^2+\frac{\kappa}{a^2}=\frac{(\sigma+\rho)^2}{\ell^2\sigma_0^2}-\frac{1}{\ell^2} 
+\frac{\mu\ell^2}{a^4}.
\label{eq031}
\end{equation}
Employing the RSII fine-tuning condition $\sigma=\sigma_0$ and (\ref{eq0001}),  Eq.\ (\ref{eq031}) may  be  expressed in the form
\begin{equation}
H_{\rm RSII}^2+\frac{\kappa}{a^2}=\frac{8\pi G_{\rm N}}{3} \rho +\frac{\rho^2}{\ell^2\sigma_0^2}
+\frac{\mu\ell^2}{a^4} ,
\label{eq022}
\end{equation}
which differs from the standard  Friedmann equation by  
the last two terms on the right-hand side.
Clearly, both versions of the RSII model yield identical brane cosmologies.

The second Friedmann equation is obtained by combining the time derivative of (\ref{eq022})
with respect to the synchronous time $\tilde{t}=\int n dt$
with the energy conservation 
\begin{equation}
\frac{d\rho}{d\tilde{t}}+3H_{\rm RSII}(\rho+p)=0.
 \label{3201}
\end{equation}
One finds
\begin{equation}
\frac{dH_{\rm RSII}}{d\tilde{t}}-\frac{\kappa}{a^2}=-4\pi G_{\rm N}(\rho+p)
 -\frac{3\rho}{\ell^2\sigma_0^2}(\rho+p)-2\frac{\mu\ell^2}{a^4} ,
 \label{eq3222}
\end{equation}
which may also be written in the form
\begin{equation}
 \frac{1}{a}\frac{d^2a}{d\tilde{t}^2}+ H_{\rm RSII}^2+\frac{\kappa}{a^2}=
 \frac{4\pi G_{\rm N}}{3}(\rho-3p)
 -\frac{\rho}{\ell^2\sigma_0^2}(\rho+3p) .
  \label{eq3226}
\end{equation}

Next we derive explicit expressions for the coordinate transformation (\ref{eq204})
for the brane position at $z=z_{\rm br}$.
Using the total differentials
\begin{equation}
 dt=\dot{t}d\tau+t'dz, \quad
 dr=\dot{r}d\tau+r'dz, 
 \label{eq105}
\end{equation}
where the prime $'$ denotes the derivative with respect to $z$,
the line element (\ref{eq3202})  transforms into
\begin{eqnarray}
ds^2&=&\left( f \dot{t}^2 -\frac{1}{f}\dot{r}^2\right)d\tau^2
-\left(\frac{1}{f}r'^2-f t'^2\right) dz^2
\nonumber \\
&&
+2\left(f\dot{t}t' -\frac{1}{f}\dot{r}r'\right)dtdz
-r^2 d\Omega_\kappa^2.
 \label{eq106}
\end{eqnarray}
The function $f$ defined in (\ref{eq101}) has the argument $r=r(\tau,z)$.
Comparing (\ref{eq106}) with (\ref{eq102}) we find 
\begin{equation}
\frac{\ell^2}{z^2}{\mathcal{N}}^2= f \dot{t}^2 -\frac{\ell^2}{z^2}\frac{\dot{\mathcal{A}}^2}{f} ,
 \label{eq107}
\end{equation}
and requiring that the off-diagonal component of the metric  vanishes  and that the $zz$ component
equals $-\ell^2/z^2$ we obtain
\begin{equation}
 t'=\frac{\dot{a}}{\dot{t}}\frac{r'}{f^2}
 \label{eq108}
\end{equation}
and 
\begin{equation}
 {\mathcal{N}}(\tau,z)=\pm \frac{f\dot{t}}{r'}= \pm \frac{l}{z}\frac{\dot{\mathcal{A}}}{ft'}.
 \label{eq110}
\end{equation}

Next we specify $z=z_{\rm br}$.
 With the help of (\ref{eq104}), (\ref{eq1041}), (\ref{eq108}), and (\ref{eq110}) we find the explicit expressions 
for $r'$, $t'$: 
\begin{equation}
 r'(\tau,z_{\rm br})= -\frac{\ell}{z_{\rm br}} \frac{f}{n}=-\frac{\ell}{z_{\rm br}}\frac{f^{3/2}}{(f^2-(\partial_t a)^2)^{1/2}},
 \label{eq111}
\end{equation}
\begin{equation}
 t'(\tau,z_{\rm br})= \frac{r'}{f^2}\frac{\dot{a}}{\dot{t}}
 =-\frac{\ell}{z_{\rm br}}\frac{f^{-1/2}}{(f^2-(\partial_t a)^2)^{1/2}}\partial_t a ,
 \label{eq112}
\end{equation}
where  the argument of $f$ is $a(t(\tau,z_{\rm br}))$,
whereas 
\begin{equation}
\dot{r}(\tau,z_{\rm br})=\dot{t}\partial_t a 
 \label{eq115}
\end{equation}
and $\dot{t}(\tau,z_{\rm br})$ remains an arbitrary function of $\tau$.
However,
the induced metric at $z=z_{\rm br}$ will have the form  ({\ref{eq028})
with $t$ replaced by $\tau$, if we identify
\begin{equation}
 \frac{\ell^2}{z_{\rm br}}\mathcal{A}^2(\tau,z_{\rm br})= a^2(\tau), 
 \quad \frac{\ell^2}{z_{\rm br}}\mathcal{N}^2(\tau,z_{\rm br})=   n^2(\tau). 
 \label{eq207}
\end{equation}
Then, from (\ref{eq1041}) and (\ref{eq207}) it follows $|\dot{t}(\tau,z_{\rm br})|=1$
yielding
\begin{equation}
t(\tau,z_{\rm br})=\pm \tau + {\rm const.} .
 \label{eq2007}
\end{equation}
Imposing that $t$ and $\tau$ increase simultaneously \cite{tetradis}
we have
\begin{equation}
 \dot{t}(\tau,z_{\rm br})=1 \quad \dot{r} =\partial_t a .
 \label{eq113}
\end{equation}

The sign  in (\ref{eq111})
is fixed from the relation between  $r'$ and the fifth component of the unit normal to the brane
in $(t,r)$ coordinates,
i.e.,
\begin{equation}
 r'(\tau,z_{\rm br})=\frac{\ell}{z_{\rm br}} n^r.
 \label{eq114}
\end{equation}
This equation follows from the transformation of $n^a=(0,0,0,0,-\ell/z_{\rm br})$ in $(\tau,z)$  to 
$n^a=(n^t,0,0,0,n^r)$  in $(t,r)$ coordinates.
Thus, with the minus  sign in (\ref{eq111}) we maintain consistency  with Eqs. (\ref{eq014}) and (\ref{eq0141}),
and the convention that $n^a$ points toward increasing $z$ (decreasing $r$).

\section{RSII/CFT connection}
\label{connection}

Here we demonstrate a connection between The RSII model and AdS/CFT correspondence.
Our derivation follows Hawking, Hertog,  and Reall \cite{hawking2} (see also Ref. \cite{tanahashi}).
We start from the  bulk action (\ref{eq001}) 
and regularize the action by placing the RSII brane near the AdS boundary, i.e., at $z=\epsilon\ell$,  $\epsilon \ll 1$  
so that the induced metric is
$h_{\mu\nu}= 1/\epsilon^2 (g^{(0)}_{\mu\nu}+\epsilon^2 \ell^2 g^{(2)}_{\mu\nu}+\ldots)$.
The bulk splits in two regions: $0\leq z\leq \epsilon\ell$  and $\epsilon\ell \leq z <\infty$, so the bulk action will
consist of two pieces. We can either discard the region $0\leq z\leq\epsilon\ell$ (one-sided regularization)
or invoke the $Z_2$ symmetry and identify two regions (two-sided regularization).
Then the regularized bulk action may be written as
\begin{equation}
S_{\rm bulk}^{\rm reg} =\gamma S_0,
 \label{eq3030}
\end{equation}
where
\begin{equation} 
S_0 =\frac{1}{8\pi G_5} \int_{z\geq \epsilon\ell} d^5x \sqrt{G} 
\left[-\frac{R^{(5)} }{2} -\Lambda _{\left(5\right)} \right] 
+S_{\rm GH}[h]
\label{eq1006} 
\end{equation}
and, as before, $\gamma=1$ for the one-sided 
and $\gamma=2$ for the two-sided regularization.
Next, we renormalize the action by adding counterterms to $S_0$
\cite{haro,hawking2}
\begin{equation} 
S_0^{\rm ren}[G]=S_0[G]+S_1[h]+S_2[h]+S_3[h],
\label{eq1007} 
\end{equation}
such that the  renormalized on-shell action is finite in the limit $\epsilon\rightarrow 0$
\begin{equation} 
S_0^{\rm ren}[g^{(0)}]=\lim_{\epsilon\rightarrow 0} S_0^{\rm ren}[h].
\label{eq1008} 
\end{equation}
The counterterms are \cite{haro}
\begin{equation} 
S_1[h]=-\frac{6}{16\pi G_5\ell}\int d^4x \sqrt{-h} , 
\label{eq4001} 
\end{equation}
\begin{equation} 
S_2[h]=-\frac{\ell}{16\pi G_5}\int d^4x \sqrt{-h}\left(-\frac{R[h]}{2} \right) ,
\label{eq4002} 
\end{equation}
\begin{eqnarray} 
S_3[h]&=&-\frac{\ell^3}{16\pi G_5}\int d^4x \sqrt{-h}\frac{\log\epsilon}{4}
 {\bigg(} R^{\mu\nu}[h]R_{\mu\nu}[h] 
 \nonumber \\
 &&
\left. 
 -\frac13 R^2[h] \right).
\label{eq4003} 
\end{eqnarray}
The last term is scheme dependent and its integrand is proportional to the 
holographic conformal anomaly
\cite{henningson}.
Now we demand that the variation with respect to $h_{\mu\nu}$ of 
the total RSII action (\ref{eq002}), which is the sum of the
regularized  bulk action (\ref{eq3030}) and 
the brane action (\ref{eq1005}), vanishes, i.e., we require
\begin{equation} 
\delta ( S_{\rm bulk}^{\rm reg}[h]+S_{\rm br}[h])=0.
\label{eq4004} 
\end{equation}
By making use of (\ref{eq4001}) this may be written as
\begin{eqnarray}
\!\!&&\delta\left[\gamma S_0^{\rm ren}- \gamma S_3-\left(\sigma-\frac{3\gamma\ell}{8\pi G_5}\right)\int\! d^4x \sqrt{-h}
\right. +
\nonumber\\
\!\!\!\!\!&&\left.
\int\! d^4x \sqrt{-h}\mathcal{L}_{\rm matt}
-\frac{\gamma \ell}{16\pi G_5}\int\! d^4x \sqrt{-h}\frac{R[h]}{2} \right]\!=\!0. 
\label{eq4005} 
\end{eqnarray}
The third term gives the contribution to the cosmological constant and may be eliminated 
by imposing the RSII fine-tuning condition (\ref{eq0012}).
The variation  of the scheme dependent $S_3$ may be combined with the first term so that
\begin{equation}
\delta (S_0^{\rm ren}-S_3)= \frac12 \int d^4x \sqrt{-h} 
\langle T^{\rm CFT}_{\mu\nu}\rangle \delta {h}^{\mu\nu} ,
\label{eq4006}
\end{equation}
where 
\begin{equation}
 \langle T^{\rm CFT}_{\mu\nu}\rangle = \frac{2}{\sqrt{-h}}
 \frac{\partial S_{\rm bulk}^{\rm (ren)}}{\partial h^{\mu\nu} }
-\frac{2}{\sqrt{-h}}
 \frac{\partial S_3}{\partial h^{\mu\nu}} .
\label{eq3004}
\end{equation}
The net effect of $\delta S_3$ is that it cancels the $\Box R$ term
in the conformal anomaly \cite{kiritsis} so  the trace of the CFT stress tensor simply reads
\begin{equation}
\langle {T^{\rm CFT}}^\mu_\mu\rangle =-\frac{\ell^3}{64\pi G_5}
 \left( R^{\mu\nu}R_{\mu\nu} -\frac13 R^2 \right).  
\label{eq3122}
\end{equation}

The variation equation (\ref{eq4005})
 yields  four-dimensional Einstein's equations on the boundary 
\begin{equation}
R_{\mu\nu}- \frac12 R g_{\mu\nu}= 8\pi G_{\rm N} (\gamma \langle T^{\rm CFT}_{\mu\nu}\rangle +T^{\rm matt}_{\mu\nu}),
 \label{eq3006}
\end{equation}
where we have employed the relation (\ref{eq0014}) to express $G_5$ in terms of Newton's constant $G_{\rm N}$. 
The quantity $T^{\rm matt}_{\mu\nu}$ is the energy-momentum tensor associated with the matter Lagrangian 
$\mathcal{L}^{\rm matt}$.
Thus, 
the dynamics of the boundary universe is governed by the energy-momentum tensor $T^{\rm CFT}_{\mu\nu}$ of
the CFT on the boundary  in addition to 
the matter energy-momentum tensor $T^{\rm matt}_{\mu\nu}$.
Obviously, the sidedness factor $\gamma$ in front of $T^{\rm CFT}_{\mu\nu}$ shows that 
the required number of copies of CFT is either one or two depending on whether the braneworld is
sitting at the cutoff boundary of a single patch of AdS$_5$  or in between two patches of AdS$_5$.
Equation (\ref{eq3006}) with (\ref{eq3106}) and $\gamma=1$ coincides with 
with Einstein's equations 
    in  Ref. \cite{haro2} derived in a different way.

From (\ref{eq3106})
with the help of (\ref{eq3120})
we obtain the vacuum expectation value of the trace of the CFT 
 energy-momentum tensor 
\begin{eqnarray}
\langle {T^{\rm CFT}}^\mu_\mu\rangle & =&{g^{(0)}}^{\mu\nu} \langle T^{\rm CFT}_{\mu\nu}\rangle 
\nonumber \\
&=&\frac{\ell^3}{16\pi G_5}
\left[  ({\rm Tr} g^{(2)})^2-{\rm Tr} (g^{(2)})^2\right].
 \label{eq3102}
\end{eqnarray}
Furthermore, using (\ref{eq3121})
we can express the trace in the form (\ref{eq3122}) 
which may be conveniently rearranged as
\begin{equation}
\langle {T^{\rm CFT}}^\mu_\mu\rangle =\frac{\ell^3}{128\pi G_5}
 \left( G_{\rm GB}-C^2\right) ,
\label{eq3123}
\end{equation}
where
\begin{equation}
G_{\rm GB}= R^{\mu\nu\rho\sigma}R_{\mu\nu\rho\sigma}-
 4 R^{\mu\nu}R_{\mu\nu} + R^2  
\label{eq3124}
\end{equation}
is the Gauss-Bonnet invariant and
\begin{equation}
C^2\equiv C^{\mu\nu\rho\sigma}C_{\mu\nu\rho\sigma}= R^{\mu\nu\rho\sigma}R_{\mu\nu\rho\sigma}
-2 R^{\mu\nu}R_{\mu\nu} + \frac13  R^2   
\label{eq3125}
\end{equation}
is the square of the Weyl tensor $C_{\mu\nu\rho\sigma}$.

The trace of the CFT energy-momentum tensor obtained in this way may be compared with the
standard conformal anomaly calculated in field theory. The general result is \cite{duff2}
\begin{equation}
\langle {T^{\rm CFT}}^\mu_\mu\rangle =
 b G_{\rm GB}-c C^2+ b'\Box R   .  
\label{eq3126}
\end{equation}
This expression will match (\ref{eq3123})
if we ignore the $\Box R $ term,  assume $b=c$,
and identify
\begin{equation}
\frac{\ell^3}{ G_5}= 128\pi c . 
\label{eq3127}
\end{equation}
For a theory with $n_{\rm s}$ scalar bosons, $n_{\rm f}$ Weyl fermions and $n_{\rm v}$ vector bosons
the standard calculations give \cite{duff2,duff}
\begin{eqnarray}
  &&b=\frac{n_{\rm s}+(11/2)n_{\rm f}+62 n_{\rm v}}{360 (4\pi)^2},
 \label{eq3119}
 \\
 && c=\frac{n_{\rm s}+3n_{\rm f}+12 n_{\rm v}}{120 (4\pi)^2}.
 \label{eq3128}
\end{eqnarray}
Hence, in general we have $b\neq c$. However
in the ${\cal{N}}=4$ U($N$) super-Yang-Mills theory, 
$n_{\rm s}=6N^2$, $n_{\rm f}=4N^2$, and $n_{\rm v}=N^2$, 
in which case the equality $b=c$ holds
and the conformal anomaly is correctly reproduced by the
holographic expression (\ref{eq3123}).
In this case 
Eq.\ (\ref{eq3127})
reads \cite{henningson}
\begin{equation}
  \frac{\ell^3}{G_5}=\frac{2N^2}{\pi}.
 \label{eq3118}
\end{equation}

It is worth mentioning  that the coefficient $c$ appears in the lowest order
quantum correction to the Newtonian potential.  The calculations based on one-loop
corrections to the graviton propagator \cite{duff3} yield the result
\begin{equation} 
\Phi(r)=\frac{G_{\rm N} M}{r}\left( 1+ \gamma\frac{128\pi c G_{\rm N}}{3r^2} \right),
\label{eq0020} 
\end{equation}
which can be compared with (\ref{eq0018}).
Here $\gamma$ is the number of copies of CFT coupled to gravity.
Applying Eq.\ (\ref{eq3127})  one finds the coefficient of the $1/r^2$ term equal to
$\gamma l^3G_{\rm N}/3 G_5$ which agrees with (\ref{eq0018}) 
if one uses the  RSII relation (\ref{eq0001}). 
Hence, as mentioned in Sec.\ \ref{introduction}, the two-sided RSII model requires 
two copies of CFT coupled to gravity.

\end{document}